\documentclass[twocolumn,nolinenumbers,trackchanges]{aastex701}
\usepackage{amsmath}
\usepackage{amssymb}
\usepackage{amsthm}
\usepackage{natbib,aasdefs,url,bm}
\usepackage{array}
\usepackage{float}
\usepackage{graphicx}
\usepackage{subfigure}
\usepackage[dvipsnames]{xcolor}
\hypersetup{linkcolor=red,citecolor=NavyBlue,filecolor=cyan,urlcolor=purple}

\def\be{\begin{equation}}
\def\ee{\end{equation}}
\def\ba{\begin{eqnarray}}
\def\ea{\end{eqnarray}}

\def\aap{\ {A\&A}\ }

\def\apj{\ {ApJ}\ }
\def\apjl{\ {ApJL}\ }
\def\apjs{\ {ApJS}\ }

\accepted{\apj}
\shorttitle{Neutrino and EM signatures from SLSN\MakeLowercase{e}}
\shortauthors{Mukhopadhyay et al.}
\begin{document}
\title{Neutrino and electromagnetic signatures from Superluminous Supernovae: a case study for SN 2017egm}
\correspondingauthor{Mainak Mukhopadhyay}
\email{mainak@fnal.gov}
\author[0000-0002-2109-5315]{Mainak Mukhopadhyay}
\affiliation{Astrophysics Theory Department, Theory Division, Fermi National Accelerator Laboratory, Batavia, Illinois 60510, USA}
\altaffiliation{FERMILAB-PUB-26-0187-T}
\affiliation{Kavli Institute for Cosmological Physics, University of Chicago, Chicago, Illinois 60637, USA}
\email{mainak@fnal.gov}
\author[0000-0003-2579-7266]{Shigeo S. Kimura}
\affiliation{Frontier Research Institute for Interdisciplinary Sciences; Astronomical Institute, Graduate School of Science, Tohoku University, Sendai 980-8578, Japan}
\email{shigeo@astr.tohoku.ac.jp}
\author[0000-0003-1336-4746]{Indrek Vurm}
\affiliation{Tartu Observatory, University of Tartu, T\~oravere, 61602 Tartumaa, Estonia}
\email{indrek.vurm@gmail.com}
\author[0000-0002-4670-7509]{Brian D. Metzger}
\affiliation{Department of Physics and Columbia Astrophysics Laboratory, Columbia University, New York, NY 10027, USA}
\affiliation{Center for Computational Astrophysics, Flatiron Institute, 162 5th Ave, New York, NY 10010, USA}
\email{bdm@columbia.edu}
\begin{abstract}
Superluminous supernovae (SLSNe) are rare transients that are $\sim 10 - 100$ times more luminous than ordinary stellar explosions, reaching peak optical luminosities $\sim 10^{44} - 10^{45}$ erg s$^{-1}$. The energy source powering SLSNe remains uncertain. In this work, we explore the multi-wavelength and multi-messenger signatures of the scenario in which SLSNe are powered by a newly born millisecond magnetar. We model the dynamical evolution and emission from the coupled system comprised of the magnetar, wind, nebula, and supernova ejecta, consistently evaluating the pair multiplicity of the wind and nebula regions, and the bulk wind Lorentz factor governing the $e^+ - e^-$ injection spectra in the nebula. We compute the thermal and non-thermal electromagnetic signatures, neutrino signatures, and investigate their detection prospects. For SN 2017egm, the nearest observed SLSNe, our prediction for high-energy gamma rays matches the recent detection by Fermi LAT. For neutrinos, using SN 2017egm a canonical SLSNe, we find that in the era of the Vera C. Rubin Observatory, a stacking analysis with upcoming neutrino observatories can lead to $3\sigma$ detection significance of neutrino events from a population of SLSNe within a decade of operation.
\end{abstract}
\section{Introduction}
The Vera C. Rubin Observatory Legacy Survey of Space and Time (LSST)~\citep{LSST:2008ijt,Blum:2022dxi} will detect roughly $\mathcal{O}(10^4)$ superluminous supernovae (SLSNe) every year (see ~\citealt{Moriya:2024gqt} for a review). These comparatively rare, but bright optical transients achieve peak absolute magnitudes $M_{\rm AB} \gtrsim - 21$ \citep{Gal-Yam:2018out}, making them a few $10 - 100$ times brighter than ordinary supernovae (SNe) resulting from the core-collapse of massive stars. Since the first confirmed observation of the superluminous SN2005ap~\citep{Quimby:2007tb}, roughly twenty years ago, a large sample of SLSNe have been discovered, leading to a population of these transients (see \citealt{Chen+23}, \citealt{Gomez+24} for recent compilations). Moreover, spectroscopic characterizations have revealed distinct hydrogen poor (Type I; \citealt{Quimby:2009ps}) and hydrogen-rich (Type II; \citealt{Smith:2009ce}) sub-populations for SLSNe. However, even with this growing and diverse population, and optimistic future detection prospects by the LSST, a clear theory for the origin of SLSNe remains elusive.

Several theories have been proposed for the power source behind SLSNe. One model postulates they arise from shock interaction between supernova ejecta with a surrounding dense (optically-thick) circumstellar medium (CSM), leading to reprocessed interaction-powered emission~\citep{Ofek:2006vt,Quimby:2009ps,Murase:2010cu,2011ApJ...729L...6C}, offering a particularly compelling explanation for many Type-II SLSNe.  Another possible origin for SLSNe are pair-instability SNe and their associated large yield of radioactive $^{56}$Ni \citep{2009Natur.462..624G,2009Natur.458..865G}. A newly-born millisecond magnetar, that is, the compact remnant left over following the core-collapse of the rapidly spinning star, were also proposed to power SLSNe~\citep{Kasen:2009tg,Woosley:2009tu,Dessart:2012vc,Kotera:2013yaa,Metzger:2013kia,Murase:2014bfa,Suzuki:2016gbg,Metzger:2017wdz,Margalit:2018bje,Omand:2023fii,Zhu:2024yrc}. The magnetar model is particularly relevant for H-poor Type-I SLSNe lacking obvious evidence for CSM interaction and will be the main focus of this work.

Given the uncertainties regarding what powers SLSNe, multi-messenger signatures, in particular, electromagnetic (EM) and neutrino signatures offer complementary probes of the potential existence of a central engine in a given SLSN, or for a population of SLSNe. For example, interaction- and magnetar-powered SLSNe can produce high-energy gamma-rays and neutrinos in the GeV-TeV range~\citep{Katz:2011fz,Murase:2013kda,Murase:2023chr,Murase:2016sqo,Fang&Metzger17,Fang+19,Murase:2021lro,Eftekhari:2020ynh} as a result of particle acceleration in the environment surrounding the explosion. However, until recently, dedicated point source searches and stacking studies~\citep{Renault-Tinacci:2017gon} for high-energy gamma rays using Fermi Large Area Telescope (LAT) data, did not yield a significant detection. The situation for neutrino searches also remains similar~\citep{Kheirandish:2022eox,Pitik:2023vcg,IceCube:2023esf} with no significant detection of signal events.

Our results are timely for several reasons. The recent detection of GeV emission from the nearest SLSNe SN 2017egm~\citep{Li:2024ics,Fermi-LAT:2026ots}, is consistent with existing predictions of a magnetar engine for SLSNe \citep{Vurm:2021dgo}. As we shall show, our model calibrated using magnetar and ejecta parameters matched to the optical observations of SN 2017egm, consistently reproduces the observed GeV emission by Fermi LAT (see bottom panel of Figure~\ref{fig:obs_sn2017egm}). In the future, upcoming EM surveys including Rubin LSST will build a large catalog of SLSNe spanning multiple wavelengths. It will then become possible to perform stacking searches for high-energy neutrinos using spatial, temporal, and energy dependent likelihood techniques, which constrain the physical parameters of the magnetar model.

This brings us to the central question addressed in this paper -- What are the multi-wavelength electromagnetic and neutrino observational prospects for magnetar-powered SLSNe? To answer this, we develop several novel aspects in the current work. We consistently follow the injection of particles and rotational energy from the magnetar into the surrounding environment, and then solve for the leptonic and hadronic cascades and interactions resulting in thermal and non-thermal photons and neutrinos \emph{both} in the nebula (dominated by $p\gamma$) and in the ejecta (dominated by $pp$). This allows us to investigate the observational prospects across the EM spectrum and for high-energy neutrinos with current and upcoming EM and neutrino telescopes, respectively. For the first time, we systematically evaluate the pair multiplicity in the wind and nebula regions and estimate the bulk Lorentz factor of the wind, where the latter governs the pair injection spectra in the nebula and subsequently the observational signatures.

The paper is organized as follows. In Section~\ref{sec:obs_slsne} we compare the results from our model and observational data for SN 2017egm to choose our fiducial parameters. We discuss our model of magnetar driven SLSNe including the ejecta properties in Section~\ref{sec:model}. The resulting observational neutrino and EM signatures are presented in Section~\ref{sec:obs_sig}. We conclude and discuss the implications of our work in Section~\ref{sec:disc}. Additional details are provided in Appendices~\ref{appsec:wind_neb_pairs},~\ref{appsec:photons_neb_ej},~\ref{appsec:hadronic_neb}, and~\ref{appsec:nu_stack}.
\section{Observational Motivations: SN 2017\MakeLowercase{egm}}
\label{sec:obs_slsne}
\begin{figure}
\centering
\includegraphics[width=0.5\textwidth]{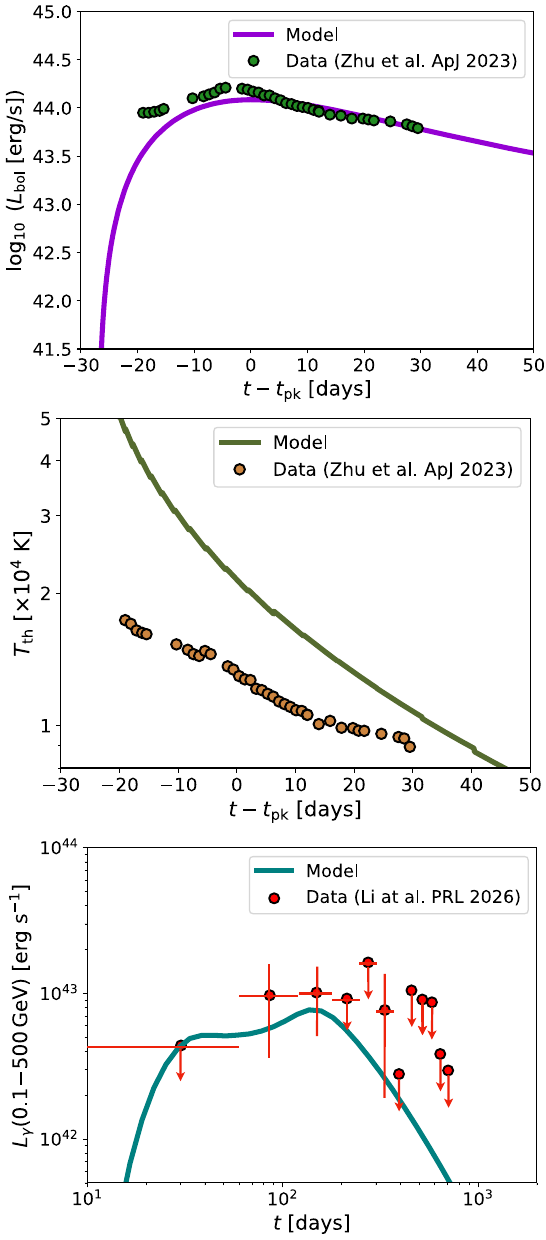}
\caption{\label{fig:obs_sn2017egm} Model predictions (solid lines) and observational data (filled circles) for SN 2017egm. The \emph{top} panel shows the time evolution of the bolometric luminosity ($L_{\rm bol}$), the \emph{middle} panel shows the evolution of the thermal temperature ($T_{\rm th}$), and the \emph{bottom} panel shows the evolution of the light curve in $0.1 - 500$ GeV band. The time at which $L_{\rm bol}$ peaks is defined as $t_{\rm pk}$ while $t$ denotes the time elapsed since the onset of the supernova.
}
\end{figure}
Over the past two decades several hundred SLSNe have been discovered~\citep{Quimby:2007tb,Gomez+24}, including several with multi-wavelength observations. Amongst this observed sample, SN 2017egm is the nearest hydrogen-poor SLSNe, with a luminosity distance $d_L \approx 135$ Mpc ($z = 0.0307$). In this section we use optical data on SN 2017egm to motivate the magnetar and ejecta parameters of the canonical model adopted in this work.

SN 2017egm was discovered on May 23, 2017, by the Gaia satellite (Gaia 17biu; \citealt{2017TNSTR.591....1D}). Radio~\citep{Bose:2017cgk} and X-ray observations (0.5 - 10 keV bands)~\citep{SWIFT:2005ngz} did not result in detections but instead placed constraining upper limits on the emission escaping the system at these energies. Additionally, recent observations have also confirmed significant GeV emission~\citep{Li:2024ics,Fermi-LAT:2026ots,Crnogorcevic:2026alv}. Most of the radiation from SLSNe occurs in the UV/optical bands, with \citet{Zhu:2023ntt} presenting a comprehensive photometric and spectroscopic survey of SN 2017egm. From this reference we obtain the the observed bolometric luminosity ($L_{\rm bol} = E_{\rm th}/t_{\rm diff}^{\rm ej}$, where $E_{\rm th}$ is the energy in thermal photons and $t_{\rm diff}^{\rm ej}$ is the diffusion time of the photons through the ejecta) and thermal temperature ($T_{\rm th}$). The observed data for $L_{\rm bol}$ (dark green circles) and $T_{\rm th}$ (dark yellow circles) is shown in Figure~\ref{fig:obs_sn2017egm} in the \emph{top} and \emph{middle} panels respectively. Predictions for $L_{\rm bol}$ and $T_{\rm th}$ from the magnetar model described in this work, are shown as light purple and green lines in the \emph{top} and \emph{middle} panels of Figure~\ref{fig:obs_sn2017egm}, respectively.

The bolometric luminosity of the model matches reasonably well with the light-curve data within $t-t_{\rm pk} \sim 50$ days, where the optical peak occurs at $t = t_{\rm pk}$. The thermal temperature evolution of the model differs most significantly (by a factor $\gtrsim 2$) at early times $t \lesssim t_{\rm pk}$; however, such differences are to be expected because we employ a simplified one-zone model for the ejecta and that does not accurately track the location of its photosphere. For purposes of motivating the ejecta and magnetar properties needed to fit SN 2017egm, this level of agreement is sufficient. Remarkably, we shall show that our model with these same chosen parameters, fit to the evolution of $L_{\rm bol}$ and $T_{\rm th}$, predict GeV emission consistent with that observed from SN 2017egm by Fermi LAT~\citep{Li:2024ics}, as shown in the \emph{bottom} panel of Figure~\ref{fig:obs_sn2017egm}.
\section{Model}
\label{sec:model}
Our model for SLSNe is similar to that developed and described in detail in~\cite{Mukhopadhyay:2024ehs, Mukhopadhyay:2025tvz} for long-lived magnetar remnants from binary neutron star mergers. In this section we shall therefore focus on the most relevant aspects of the model and highlight those parts that differ from our previous works.

The core-collapse of a rapidly spinning star leaves behind a remnant neutron star characterized by a millisecond initial spin period ($P_i \sim 1 - 5$ ms) and a strong dipolar magnetic field of strength $B_d \sim 10^{13} - 10^{15}$ G~\citep{Gomez+24,Nicholl_2017}. This remnant is defined as a \emph{magnetar}. The magnetar loses its spin energy due to magnetically-driven (pulsar-like) winds. This energy resulting from the spinning down of the pulsar, also known as the spindown energy, is deposited behind an expanding ejecta and leads to the formation of a hot nebula. The magnetar's spindown energy acts as the reservoir for the thermal ($E_{\rm th}$), non-thermal ($E_{\rm nth}$), magnetic ($E_B$), and kinetic energy via PdV work done on the system. The dynamics of the system is solved by evolving these energies~\citep{Mukhopadhyay:2024ehs, Mukhopadhyay:2025tvz,Mukhopadhyay:2026wrv} using the following equations (see also \citealt{Metzger&Piro14})
\begin{equation}
\label{eq:nth}
\frac{d E_{\rm nth}}{dt} = L_{\rm sd} - \frac{E_{\rm nth}}{R} \frac{d R}{dt} - \frac{E_{\rm nth}}{t^{\rm neb}_{\rm diff}}\,,
\end{equation}
\begin{equation}
\label{eq:th}
\frac{d E_{\rm th}}{dt} = \big( 1 - \mathcal{A} \big) \frac{E_{\rm nth}}{t^{\rm neb}_{\rm diff}} - \frac{E_{\rm th}}{R} \frac{d R}{dt} - \frac{E_{\rm th}}{t^{\rm ej}_{\rm diff}} + Q^{\rm heat}_{\rm Ni}\,,
\end{equation}
\begin{equation}
\label{eq:eb}
\frac{d E_{B}}{dt} = \varepsilon_B L_{\rm sd} - \frac{E_B}{R} \frac{dR}{dt}\,,
\end{equation}
where $L_{\rm sd}$ is the spindown luminosity of the magnetar, $R$ is the radial distance of the nebula-ejecta boundary from the center of the magnetar, $t_{\rm diff}^{\rm neb}$ and $t_{\rm diff}^{\rm ej}$ are the photon diffusion timescales in the nebula and in the ejecta, respectively, $\mathcal{A}$ is the fraction of non-thermal photons that escape from the ejecta, $\varepsilon_B$ is a parameter that determines the magnetic field strength in the nebula, and $Q_{\rm heat}^{\rm Ni}$ is the rate of heating from the radioactive decay of $^{56}$Ni, discussed in Appendix~\ref{appsubsec:ni_heat}. The non-thermal energy in photons is proportional to the spindown luminosity which acts as the source term. The loss terms include the $PdV$ work done by the non-thermal photons on the ejecta and their diffusion through the nebula-ejecta boundary, given by the second and third terms respectively in Equation~\eqref{eq:nth}. The energy in thermal photons is sourced from the reprocessing of a fraction $(1-\mathcal{A})$ of the non-thermal photons and from the decay of $^{56}$Ni (first and fourth terms respectively in Eq.~\ref{eq:th}). Similar to the non-thermal photons the $PdV$ work by the thermal photons and their diffusion from the ejecta act as loss terms in this case. Finally, the the magnetic energy evolves proportional to $L_{\rm sd}$ and also does $PdV$ work which acts as the loss term.

\begin{table*}
\centering
\caption{Model parameters}
\label{tab:params}
\begin{center}
Stellar progenitor, supernova ejecta, and magnetar properties
\end{center}
\begin{minipage}{\linewidth}
\centering
\begin{center}
\begin{tabular}{ccccccccccc}
\hline
\hline
$M_i$ & $M_f$ & $M_*$ & $R_*$ & $I$ & $M_{\rm ej}$ & $E_{\rm kin}$ & $P_i$ & $B_d$ & $v_{\rm ej}$ & $\kappa_{\rm ej}$\\
\hline
$16\ M_\odot$  & $3.65\ M_\odot$ & $1.42\ M_\odot$ & $10^{6}$\ \rm cm & $10^{45}\ {\rm g\ cm}^2$ &  $2.23\ M_\odot$ & $5.08 \times 10^{51}\ \rm erg$ & $4\ \rm ms$ & $5 \times 10^{13}$ G & $0.05\ c$ & $0.1\ {\rm cm}^{2}{\rm g}^{-1}$\\
\hline
\end{tabular}
\end{center}
\end{minipage}
\begin{center}
Elemental ejecta composition from Model \emph{3p65Dx2} of~\citet{Dessart:2016fun}
\end{center}
\begin{minipage}{\textwidth}
\centering
\begin{tabular}{cccccccccccc}
\hline
\hline
Element & H & He & C  & N  & O  & Si & S  & Ca & Fe & Co & $^{56}$Ni \\
\hline
Comp. ($M_\odot \times 10^{-2}$) & $0.464$ & $155$ & $9.33$ & $1.07$ & $25.1$ & $7.64$ & $2.86$ & $0.548$ & $0.360$ & $0$ & $10.5$ \\
\hline
Num Frac. ($N_0$) &  $0.0109$ &  $0.9184$ & $0.0184$ & $0.0018$ & $0.0372$ &  $0.0064$ & $0.0021$ & $0.0003$ & $0.0002$ & $0.0$ & $0.0042$ \\
\hline
\end{tabular}
\end{minipage}
\end{table*}
\begin{figure}
\centering
\includegraphics[width=0.49\textwidth]{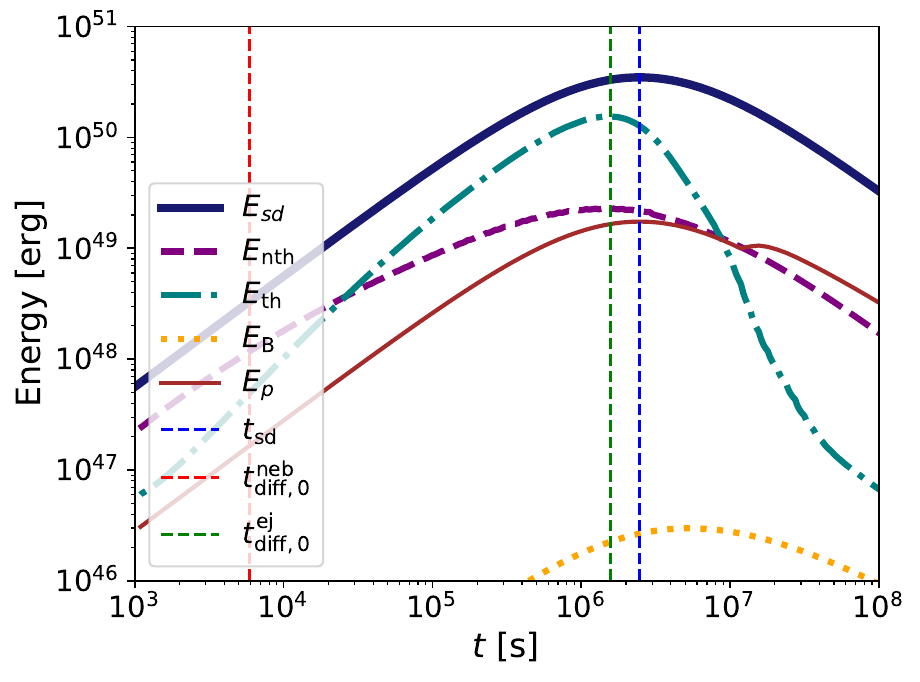}
\caption{\label{fig:energies} Time evolution of various components of the energy along with spin down and characteristic nebula and ejecta diffusion timescales, for our fiducial magnetar model for SN2017egm.
}
\end{figure}
For the ejecta mass $M_{\rm ej}$, kinetic energy $E_{\rm kin}$, and elemental composition we use model the \emph{3p65Dx2} of~\citet[][see Table B2 there and also previous work in~\citealt{Dessart:2015mga}]{Dessart:2016fun}. The initial velocity of the ejecta is given by $v_{\rm ej} = \sqrt{2 E_{\rm kin}/M_{\rm ej}}$. Given these ejecta properties, we scanned over the dipole field strength ($B_d$) and initial spin period $P_i$ of the magnetar, finding that the combination $\{B_d = 5 \times 10^{13}$ G, $P_i = 4$ ms$\}$ results in an optical light curve most closely matching the $L_{\rm bol}(t)$ and $T_{\rm th}(t)$ evolution of SN 2017egm (Figure~\ref{fig:obs_sn2017egm}).  The parameters of the progenitor star, ejecta, and magnetar remnant are summarized in Table~\ref{tab:params}. 

Given these parameters, the evolution of the various energies, obtained from solving Equations~\eqref{eq:nth},~\eqref{eq:th}, and~\eqref{eq:eb},  are shown in Figure~\ref{fig:energies}. The spindown energy peaks around the spindown timescale ($t_{\rm sd}$), defined as the ratio between the initial rotational energy of the magnetar and the initial spindown luminosity $L_{\rm sd}^{\rm init}(P_i,B_d)$. For the fiducial parameter set, we have $t_{\rm sd} \sim 2.5 \times 10^6$ s and the peak of $E_{\rm sd} \sim 4 \times 10^{50}$ erg. The non-thermal energy, $E_{\rm nth} \propto E_{\rm sd}$ and hence closely follows the evolution of $E_{\rm sd}$ at initial times. However, as diffusion from the nebula starts becoming relevant after the characteristic timescale $t_{\rm diff,0}^{\rm neb} \sim 5.9 \times 10^3$ s, non-thermal photons begin to diffuse out of the nebula through the ejecta, which leads to the evolution of $E_{\rm nth}$ to deviate from $E_{\rm sd}$. The thermal energy evolution dominates the non-thermal component for $t \gtrsim 2 \times 10^4$ s. This can be understood by the fact that the source term for the thermal photons is the reprocessed component of the non-thermal photons (Equation~~\ref{eq:th}), where the fraction of reprocessed component is given by $(1 - \mathcal{A})$, $\mathcal{A}$ being the ejecta albedo. This albedo is systematically evaluated based on the ejecta composition and its ionization state\footnote{We assume that the ejecta elements are singly ionized, which is a reasonable approximation given that the most abundant element in the ejecta, He, is largely neutral~\citep{Ekanger:2026coc}. The ionization state most prominently affects the free-free absorption and hence the late time radio emission.}. For $t\lesssim 10^6$ s, $\mathcal{A} \sim 0$ and hence most of the non-thermal photons are reprocessed into thermal photons. However, given the ejecta mass and composition, they are opaque to thermal photons for $t<t_{\rm diff,0}^{\rm ej} \sim t_{\rm sd}$, when most of the spindown energy is deposited, leading to $E_{\rm th}$ dominating over $E_{\rm nth}$. At late times $t>t_{\rm diff,0}^{\rm ej} \sim 1.6 \times 10^6$ s not only does $E_{\rm sd}$ decrease, but also diffusion of thermal photons becomes significant and hence $E_{\rm th}$ decreases rapidly.

The magnetic energy in the nebula $E_B \propto E_{\rm sd}$ and therefore closely follows the injected spin-down energy $E_{\rm sd}$ in proportion to the nebula magnetization, $\varepsilon_B$.  Although the magnetar wind on small scales is dominated by Poynting flux and extremely highly magnetized, various dissipation processes can occur in the wind and nebula that may act to reduce the magnetic field at the expense of plasma heating.  Following model fits to optical light curve of 2017egm \citep{Nicholl:2017mnb,Nicholl:2018cam} by \citet{Vurm:2021dgo}, we take $\varepsilon_B \sim 10^{-4}$.   We discuss the energy placed into relativistic protons $E_p$ in Section~\ref{subsec:nu_prod}.

\subsection{Electron and photon spectra}
\label{subsec:pairs_photons}
Here we discuss the pair injection spectra, the resulting photons, and their attenuations in our model. The magnetar wind is composed of pairs extracted from the magnetosphere as a result of $B - \gamma$ and $\gamma - \gamma$ processes. At initial times, the bulk Lorentz factor of the wind $\Gamma_w \gg 1$, that is, it is in the ultra-relativistic regime. However, its encounter with the expanding ejecta decelerates the wind, which leads to dissipation of their kinetic energy and formation of a wind termination shock (TS). The shocked region downstream of the TS forms the hot nebula, where the bulk kinetic energy of the wind is converted into particle energy and magnetic fields. Thus, the distribution of particles in the downstream will naturally be influenced by the characteristic energy available per particle in the upstream wind. We parametrize this characteristic energy scale by an effective wind Lorentz factor $\Gamma_w$, defined as
\begin{equation}
\Gamma_w \approx \frac{L_{\rm sd}}{4 \pi m_e c^2 n^w_{e^\pm} R_s^2 c} \,,
\end{equation}
where the rate of energy extraction by the wind $L_w \sim L_{\rm sd}$, the pair number density in the wind region is given by $n^w_{e^\pm}$, $R_s$ is the approximate radial distance of the TS from the central engine. For a cold, kinetically-dominated wind (non-magnetized), $\Gamma_w$ corresponds to the usual bulk Lorentz factor of the wind, while for the magnetized scenario it can be interpreted more generally as the available wind energy per particle, including magnetic contributions.

We assume that the pairs entering the nebula at the TS are well-described by a broken power law injection set by the break Lorentz factor of the pairs $\gamma_{e,\rm br} \approx \Gamma_w$. This is reasonable since the pairs are injected into the nebula from the wind region, with the break energy set by the characteristic wind energy available per particle. In the shock-driven reconnection picture, the relatively hard power-law index below the break can be associated with the low-energy component produced by magnetic dissipation or reconnection in the upstream wind and around the TS, whereas the softer index above the break describes the high-energy non-thermal tail produced predominantly by acceleration close to the TS~\citep{Sironi:2011zf}. Furthermore, phenomenologically such broken power-law pair injection spectra are also supported by broadband modeling of galactic pulsar wind nebulae~\citep{Torres:2014iua}. Hence, the primary pair injection spectrum has the following form
\be
Q_e^{\rm inj} (\gamma_e) = Q_e^0 \exp\bigg( -\frac{\gamma_e}{\gamma_{e,\rm cut}} \bigg)
\begin{cases}
\bigg(\frac{\gamma_e}{\gamma_{e,\rm br}}\bigg)^{-1.5} \hspace{-1.5em},\ \gamma_m \leq \gamma_e \leq \gamma_{e,\rm br}\\
\bigg(\frac{\gamma_e}{\gamma_{e,\rm br}}\bigg)^{-2.5}\hspace{-1.5em},\ \gamma_{e,\rm br}< \gamma_e \leq \gamma_M\,,
\end{cases}
\ee
where the electron Lorentz factor $\gamma_e = \varepsilon_e/(m_e c^2)$, $\varepsilon_e$ is the comoving energy of the electron, $\gamma_m$ and $\gamma_M$ are the minimum and maximum electron Lorentz factors chosen to be $1$ and $10^9$ respectively (note that this choice has no consequence on the results). The cut-off Lorentz factor $\gamma_{e,\rm cut}$ is determined self-consistently by balancing the cooling and acceleration timescales for the electrons. The normalization $Q_e^0 = L_{\rm sd}^{\rm com}/\big( m_e c^2 \int d\gamma_e\ \gamma_e\ Q_e^{\rm inj}(\gamma_e)/Q_e^0\big)$, where $L_{\rm sd}^{\rm com}$ is the comoving $L_{\rm sd}$. The pair multiplicity\footnote{Note that the pair multiplicity defined here by $Y$ is analogous to the \emph{pair yield}, that is, the fraction of injected power converted into pair rest mass, as described in~\cite{1987MNRAS.227..403S} (see Equation 2.25 there) and~\cite{1988ApJ...335..786Z} (see Equation 12 there).} in the wind and the nebula can be defined as $Y = \dot{N}_{e^\pm}^{\rm c} m_e c^2/L_{\rm sd}$, where $\dot{N}_{e^\pm}^{\rm c}$ is the pair creation rate in the wind or nebular regions. This is important in determining the diffusion timescale of the photons in the nebula (see Equation~\ref{eq:nth}) $t_{\rm diff}^{\rm neb} \approx (R/c) (1 + \tau_{\rm diff}^{\rm neb})$, where the Thomson optical depth is given by $\tau_{\rm diff}^{\rm neb} \approx \big(4 Y L_{\rm sd} \sigma_T/(\pi m_e c^3 R \Gamma_{\rm ej}^2)\big)^{1/2}$. Appendices~\ref{appsubsec:pairs} and~\ref{appsubsec:wind} provide a detailed calculation of the pair multiplicity and the bulk wind Lorentz factor, respectively.

The injected pairs are transported and the quasi-steady state spectrum is obtained by considering synchrotron, inverse Compton (IC), and dynamical losses. The energy distribution of the non-thermal photons are then computed through the full electromagnetic cascade via synchrotron, IC, and $\gamma \gamma$ processes. The energy density of the thermal photons can be evaluated using the thermal temperature $T_{\rm th} \propto E_{\rm th}^{1/4}$. IC scattering of thermal photons which have a large energy density dominates the non-thermal photon spectra, while synchrotron is significant for $E_\gamma \lesssim 1$ eV. At low energies ($E_\gamma \lesssim 1$ eV), synchrotron self-absorption (SSA) effects become important and at high energies ($E_\gamma \gtrsim 10$ GeV) Klein-Nishina and $\gamma \gamma$ pair production suppress the energy density in non-thermal photons. The latter is also taken into account consistently in our model where we compute the secondary pairs produced from $\gamma \gamma$ pair production processes to account for the non-thermal photons. The resulting photon spectra in the nebula is discussed in Appendix~\ref{appsubsec:neb_photons} and will be relevant for providing the radiation field for neutrino production discussed in Section~\ref{subsec:nu_prod}.

The nebular photons further suffer attenuation in the supernova ejecta via free-free, bound-bound, bound-free, Compton, and Bethe-Heitler processes in the radio, IR/Optical/UV, UV/X-ray, MeV, GeV-TeV wavelengths. We present the relevant details about attenuation across various wavelengths and its evolution with time in Appendix~\ref{appsubsec:attn}.

\subsection{Neutrino production}
\label{subsec:nu_prod}
In this subsection, we discuss the proton injection spectrum, acceleration of the injected CR protons, and their subsequent interactions resulting in neutrino production from magnetar-powered SLSNe. We consider neutrino production in the nebular and the ejecta regions. The former is dominated by photohadronic ($p\gamma$) interactions, while the latter by hadronuclear ($pp$) interactions owing to the dense ejecta.

The plasma within the magnetosphere of the pulsar is dragged along the magnetic field lines and hence co-rotates with the magnetar. This co-rotation of charged particles creates an electric field, a component of which is not perpendicular to the magnetic field. This electric field can be used to extract charges from the surface of the magnetar. For simplicity, we restrict ourselves to protons\footnote{For a discussion on ions or nuclei see discussion in~\cite{Mukhopadhyay:2024ehs}.}. The maximum charge density that can be extracted can be estimated using the Goldreich-Julian charge density~\citep{Goldreich:1969sb} such that the rate of CR proton production is given by $\dot{N}_p = \big( (4 \pi^2/e c) (R_*^3 B_d/P^2)\big)$, where the spin period evolution is governed by $P(t) = P_i \big( 1 + t/t_{\rm sd} \big)^{1/2}$.

The polar cap region of the magnetar is capable of accelerating protons in the potential gap. Thus, the injection spectra $d \dot{N}_{p}^{\rm inj}/d\varepsilon^\prime_p = \dot{N}_p \delta \big( \varepsilon_p^\prime - \varepsilon_p^{\prime\rm cutoff,pc} \big)$, where $\varepsilon_p^{\prime\rm cutoff,pc}$ is the cut-off energy in the polar cap region. The primed quantities denote the comoving quantities unless otherwise stated. The Dirac-delta function for the injection spectra~\citep{Arons:2002yj} is motivated by the fact that the potential gap created by the electric field is shorted at periodic intervals and therefore the CR protons are accelerated at discrete energies. The cut-off energy for the protons in the polar cap region (which determine their injection spectra) is obtained by comparing the maximum energy that the protons can gain while they are accelerated across the potential gap and the energy loss due to curvature radiation, such that at any instant the minimum energy of the two is chosen as $\varepsilon_p^{\prime\rm cutoff,pc}$.

The CR protons can also be re-accelerated at the TS as a result of shocks or magnetic reconnection~\citep{Sironi:2011zf,Sironi:2014jfa}. The cutoff in the TS region $\varepsilon_p^{\prime\rm cutoff,TS}$ is obtained by balancing the acceleration and cooling timescales for the protons. This finally gives the maximum energy to which the CR protons can be accelerated as $\varepsilon_p^{\prime\rm max} = \min [\varepsilon_p^{\prime\rm cutoff,pc}, \varepsilon_p^{\prime\rm cutoff,TS}]$. The evolution of $\varepsilon_p^{\prime\rm cutoff,pc}$, $\varepsilon_p^{\prime\rm cutoff,TS}$, and $\varepsilon_p^{\prime\rm max}$ is discussed in Appendix~\ref{appsubsec:proton_energies}. The total energy in protons which can be estimated using $E_p(t) = \dot{N}_{p,\rm inj} \varepsilon_p^{\prime\rm max} t$ peaks around $t_{\rm sd}$ and is shown as a brown line in Figure~\ref{fig:energies}.

The quasi-steady state energy spectrum for the accelerated CR protons in the nebula ($d N_{p}/d\varepsilon^\prime_p$) is obtained by solving the transport equation~\citep{Kimura:2019yjo} taking into account cooling and escape. The CR protons then interact with the photons or protons in the nebula and ejecta to produce charged and neutral pions amongst other decay products. A fraction of the charged pions produce neutrinos while the neutral pions produce photons. Since the nebula is filled with non-thermal and thermal photons, the CR protons mostly cool through the photomeson channel, while the ejecta consists of a large number of protons and hence $pp$ interactions serve as the dominant cooling channel. The relevant acceleration and cooling timescales for the CR protons are discussed in Appendix~\ref{appsubsec:timescales}. In particular, for the proton cooling channels we take into account $pp$, $p\gamma$, proton synchrotron, Bethe-Heitler, and adiabatic processes. Besides cooling, we also consider advective escape and diffusion of protons contributing to their energy loss timescales. To compute the neutrino spectra we use the fitting formulae from~\cite{Kelner:2008ke}. The suppression factors associated with the cooling of pions and muons are also taken into account consistently. For the observed neutrino spectra, we also account for the oscillation effects during their propagation from the SLSNe to an observer on Earth.

For the $pp$ component in the ejecta, the comoving target nucleon number density can be estimated $n_p^{\prime\rm ej} = \rho^{\prime \rm ej}/m_u$, where $\rho^{\prime \rm ej}$ is the comoving density of the ejecta and $m_u$ is the atomic mass unit. The CR protons that escape the nebula serve as the source for the $pp$ interactions in the ejecta, such that the injection rate is given by $d \dot{N}_{p}^{\rm inj, ej}/d\varepsilon^\prime_p = \big(d N_{p}/d\varepsilon^\prime_p\big)/t^{\prime}_{\rm esc}$, $t^{\prime}_{\rm esc} = \rm max[R^2/D_c(\varepsilon_p^\prime), R/c]$ is the escape timescale of CR protons in the nebula, the first term accounts for diffusion~\citep{Harari:2013pea} while the second term accounts for free-streaming of the protons from the nebula to the ejecta. The $pp$ interaction rate can be estimated as $t_{pp}^{\prime -1, \rm ej} \simeq n_p^{\prime\rm ej} \sigma_{pp} \kappa_{pp} c$, where $\sigma_{pp}$ and $\kappa_{pp}$ are the proton interaction cross-section and inelasticity respectively~\citep{Kelner:2006tc}. The quasi-steady state CR proton spectrum in the ejecta is given by  $d N_{p}^{\rm ej}/d\varepsilon^\prime_p = \big(d \dot{N}_{p}^{\rm inj, ej}/d\varepsilon^\prime_p\big) t^{\prime \rm ej}_{\rm loss}$, where $t^{\prime -1, \rm ej}_{\rm loss} = t_{pp}^{-1, \rm ej} + t_{\rm exp}^{-1, \rm ej} + t_{\rm esc}^{-1, \rm ej}$. The expansion timescale of the ejecta is given by $t_{\rm exp}^{\rm ej} = R/(\Gamma_{\rm ej} v)$ and $t_{\rm esc}^{\rm ej}$ represents the escape timescale of the protons from the ejecta. Similar to the neutrino computation in the nebula, the resulting ejecta proton spectrum $d N_{p}^{\rm ej}/d\varepsilon^\prime_p$ is then used to compute the secondary gamma rays, pairs, and neutrinos. We further include synchrotron cooling of secondary pions and muons before applying flavor mixing to obtain the total all-flavor neutrino output from the ejecta.
\section{Observational signatures}
\label{sec:obs_sig}
This section focuses on the main results of this work, that is, the observational multi-wavelength EM and neutrino signatures from a canonical magnetar-powered SLSNe, like SN 2017egm, at a typical source distance $d_L  = 100$ Mpc. We present our results in the context of current and upcoming EM and neutrino telescopes and elaborate on their detection prospects.
\subsection{Electromagnetic signatures}
\label{subsec:em_sig}
\begin{figure}
\centering
\includegraphics[width=0.45\textwidth]{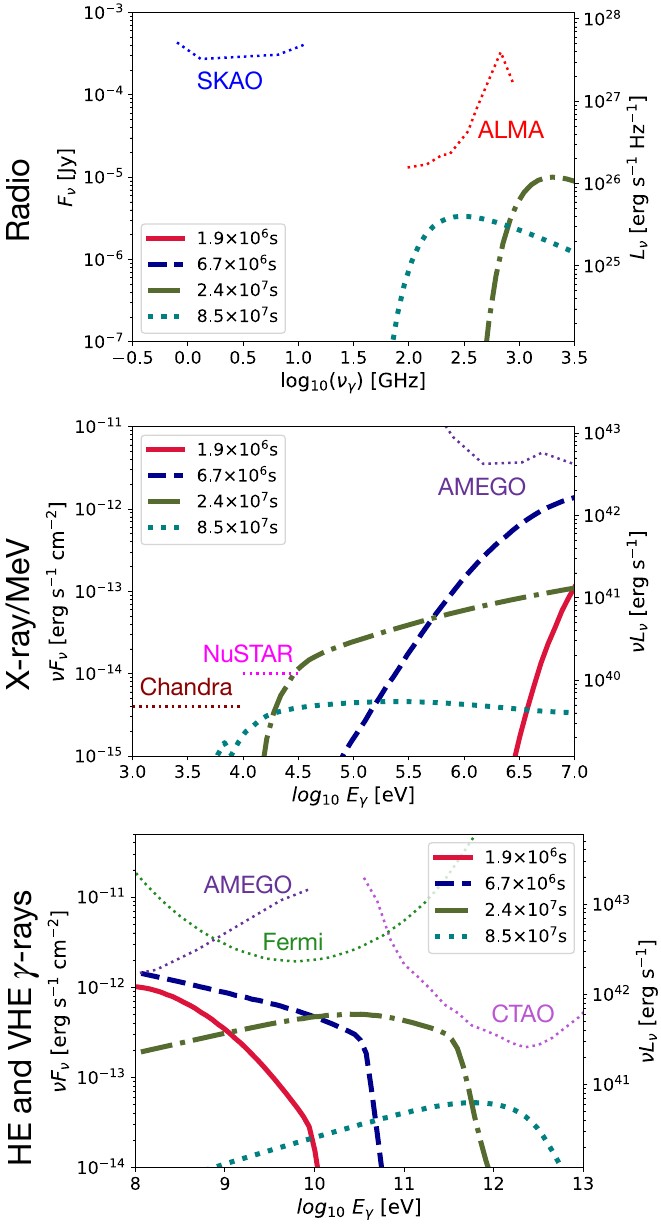}
\caption{\label{fig:emspectra}Predicted electromagnetic spectra from our fiducial magnetar-powered SLSNe model at $100$ Mpc corresponding to the radio, X-ray, MeV, and gamma-ray bands at different time snapshots. The sensitivity curves for the relevant detectors in the given EM bands are also shown. Note that $F_\nu$ denotes spectral flux density, $L_\nu$ is the spectral luminosity, and $\nu L_\nu$ is the differential luminosity. The radio spectra \emph{(top)} shows flux density while the \emph{middle} and \emph{bottom} panels show flux.
}
\end{figure}
\begin{figure}
\centering
\includegraphics[width=0.45\textwidth]{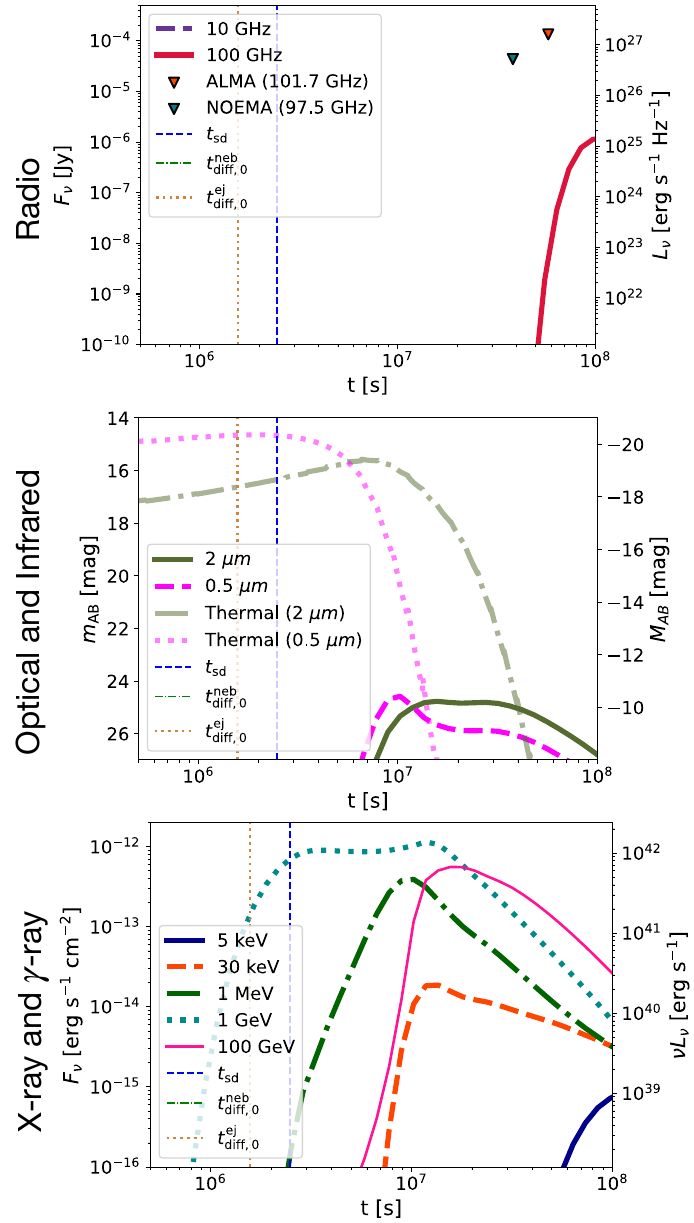}
\caption{\label{fig:lc}Light curves from a SLSNe at $100$ Mpc corresponding to the radio, optical, infrared, X-ray and gamma-ray bands. The characteristic spindown ($t_{\rm sd}$) and ejecta diffusion  ($t_{\rm diff,0}^{\rm ej}$) timescales are also shown. The representative frequencies (for radio), wavelengths (for optical and infrared), and energies (for X-ray and $\gamma-$ rays) are motivated by relevant bands from observational telescopes shown in Figure~\ref{fig:emspectra}. The $3\sigma$ upper limits from ALMA and NOEMA are also shown from~\cite{Murase:2021lro}.
The apparent and absolute magnitudes are shown by $m_{AB}$ and $M_{AB}$ respectively. 
}
\end{figure}
We divide the observational EM signatures into different wavelengths and discuss their corresponding features and detectability given the relevant telescopes in that band. The predicted photon spectra in the mm/radio, X-ray/MeV, and GeV and TeV bands for different temporal snapshots are shown in Figure~\ref{fig:emspectra}. Sensitivity curves for various current and upcoming observational facilities (see Table 1 in~\citealt{Mukhopadhyay:2025tvz}, acronyms, and references therein) are also overlaid. Sources of photon attenuation relevant for interpreting features in the EM spectra are discussed in Appendix~\ref{appsubsec:attn}. Light curves are shown in Figure~\ref{fig:lc} and correspond to radio, optical, infrared, X-ray, and gamma-ray bands. For comparison, we overlay the spindown and other key timescales as vertical lines.

The radio band spectrum is cut-off at low frequencies by SSA in the nebula and free-free absorption in the ejecta, respectively, both of which decrease monotonically in frequency with time as the system expands. Even at late times ($\sim$ a few years post-collapse) free-free absorption in the ejecta completely attenuates $\lesssim 100$ GHz emission. Millimeter/sub-mm radio telescopes like ALMA can potentially detect the late time radio emission as can be inferred from the 100 GHz radio light curve. We also show the ALMA and NOEMA $3\sigma$ upper limits from radio observations of SN 2017egm~\citep{Murase:2021lro}. Our radio light curves at $100$ GHz are indeed consistent with the upper limits. However, at these frequencies the forward shock (afterglow) generated as the ejecta collides with circumstellar material might outshine our predicted signal~\citep{Chen:2026mnf}. Moreover, the radio signal might also be enhanced at late times from pairs that accumulate in the nebula over time (e.g., \citealt{Margalit&Metzger18}), which our current model does not take into account.

The ejecta attenuation, which depends on its elemental composition, also suppresses the observed X-ray and MeV photon spectra. The observation of soft X-rays in telescopes like Chandra or XMM-Newton are thus unlikely, which can also be noted from the 5 keV X-ray light curve having a tiny ($\sim 10^{-15} {\rm erg\ s}^{-1}\rm cm^{-2}$) flux at very late times. However, hard X-rays are likely to be observed providing NuSTAR with optimistic detection prospects. For a typical NuSTAR detection band of 30 keV, the light curve reaches a peak sensitivity of $\sim {\rm a\ few\ } 10^{-14} {\rm erg\ s}^{-1}\rm cm^{-2}$ at $\sim 1$ year post collapse. In the MeV band, the peak occurs $\sim 10^7$ s post collapse corresponding to the ejecta becoming transparent to MeV wavelengths. This indicates that upcoming telescopes like AMEGO/AMEGO-X can potentially detect the source.

The dominant source of attenuation of GeV gamma-rays are by $\gamma \gamma$ pair production in the nebula, Bethe-Heitler, and Compton processes in the ejecta. However, the $\gamma \gamma$ cutoff energy shifts to higher energies with time, allowing for $\sim$ TeV emission within a year from collapse; the ejecta becomes completely transparent to gamma-rays on a similar timescale. The situation is even better for GeV gamma rays which can become observable in telescopes like Fermi roughly around spindown timescale. Since the GeV photons are effectively unattenuated around $t_{\rm sd}$, the GeV lightcurve peaks around $t \sim t_{\rm sd}$ and subsequently forms a plateau. Thus, GeV gamma ray telescopes like Fermi indeed have optimistic chances of detecting high-energy gamma-ray emission from magnetar-powered SLSNe. Thus, the observed evidence of GeV emission from SN 2017egm and the compatibility of our predictions for GeV gamma-rays is not surprising (see Figure~\ref{fig:obs_sn2017egm} \emph{bottom} panel). Moreover, at later stages very high-energy gamma-rays in the 100 GeV band might also be detected by CTAO from such a source.

The \emph{middle} panel of Figure~\ref{fig:lc} shows both the thermal and non-thermal optical light curves. As discussed in Section~\ref{sec:model}, the thermal components completely dominate over the non-thermal component for SLSNe\footnote{
Although completely overwhelmed by the thermal emission at optical wavelengths, the non-thermal component is constrained by hard X-ray and gamma-ray observations.}. We note that as expected, with a peak absolute AB magnitude of $\sim -20$, SLSNe are optically very bright. Thus, optical telescopes like Rubin LSST, will be observing $\mathcal{O}(10^5)$ SLSNe within the next decade (see Appendix~\ref{appsec:nu_stack}), assuming Rubin LSST observes 1/3 of all SLSNe. The SLSNe catalogs obtained from these surveys will then allow for stacking searches for detecting neutrino counterparts and gaining further insights into the properties of SLSNe.
\subsection{Neutrino signatures}
\begin{figure*}
\centering
\includegraphics[width=0.49\textwidth]{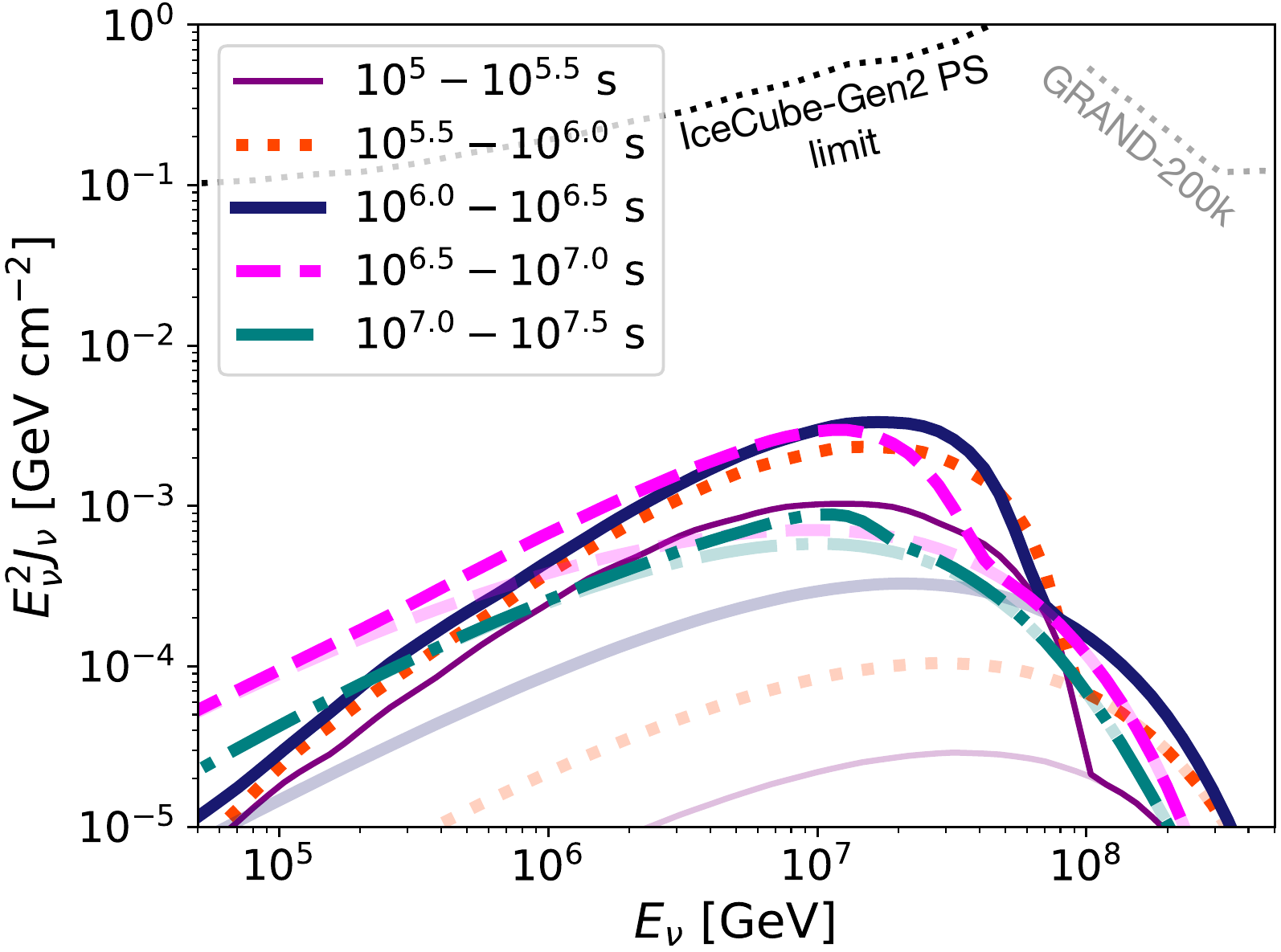}
\includegraphics[width=0.48\textwidth]{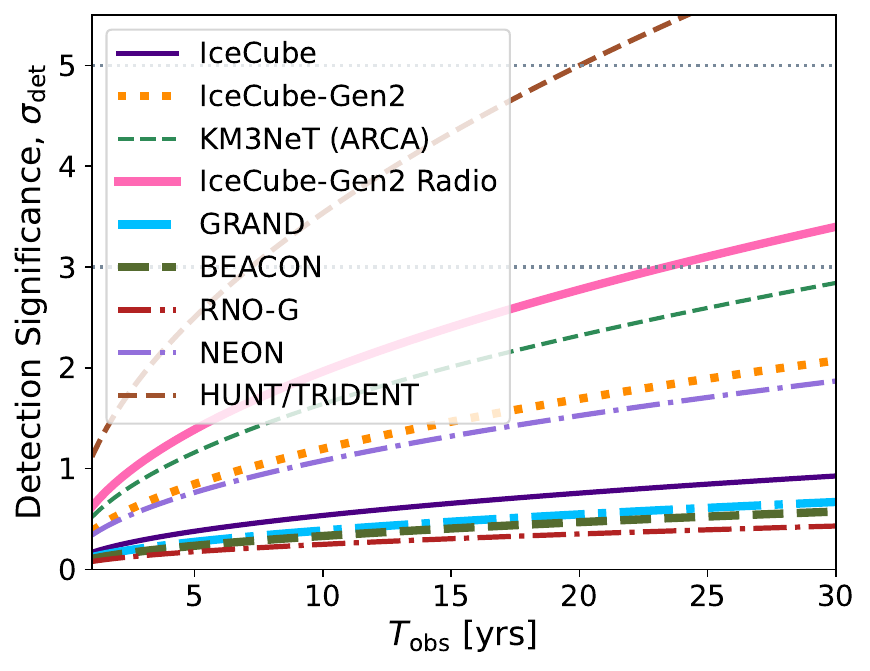}
\caption{\label{fig:nu} \emph{Left: }Observed neutrino fluence (all-flavors) for a magnetar-powered SLSN at 100 Mpc for timescales post core-collapse ranging from a few hours to a year. The corresponding $pp$ component from the ejecta is shown in fainter shades. \emph{Right: }The detection significance for various current, upcoming, and proposed neutrino telescopes given the observation time ($T_{\rm obs}$) as a result of stacking a population of observed SLSNe using Rubin LSST within redshift of $z_{\rm max} = 1$, where we assume Rubin LSST observes $1/3$ of all SLSNe. Note that the all-flavor day-averaged effective area including the appropriate field of view is used (see Appendix~\ref{appsec:nu_stack} for details) for IceCube, RNO-G, IceCube-Gen2, IceCube-Gen2 Radio, GRAND, and BEACON.
}
\end{figure*}
The observed all-flavor neutrino fluence resulting from our canonical magnetar-powered SLSNe, including both the $p\gamma$ component in the nebula and the $pp$ component in the ejecta, is shown in Figure~\ref{fig:nu} \emph{(left)}. Relevant results for maximum proton energy and timescales are shown in Appendices~\ref{appsubsec:proton_energies} and~\ref{appsubsec:timescales}, respectively. To illustrate the temporal evolution of the resulting neutrino fluence, we compute the fluence integrated over successive logarithmic time bins spanning half a decade.  Furthermore, for comparison we show the point source sensitivities of upcoming neutrino telescopes like IceCube-Gen2~\citep{IceCube-Gen2:2020qha,Gen2_TDR} and Giant Radio Array for Neutrino Detection (GRAND-200k,~\citealt{GRAND:2018iaj}).

The neutrino fluence peaks in the energy range $10^7 - 10^8$ GeV  at a fluence $\sim 3 \times 10^{-3}\ {\rm GeV cm}^{-2}$ on a timescale of a few weeks to a month, $t \sim 10^6 - 10^{6.5}$ s post collapse. This is reasonable because the total energy in protons peaks near the spindown timescale around $E_p \sim 1.7 \times 10^{49}$ erg (from Figure~\ref{fig:energies}). Thus, the peak neutrino fluence can be estimated using $E_\nu^2 J_\nu \sim \big( 1/(4 \pi d_L^2) (3/8) E_p f_{p\gamma} f_{\rm bol} f_{\pi,\rm sup}\big)$, where $f_{p\gamma}$, $f_{\rm bol}$, and $f_{\pi,\rm sup}$ are the $p\gamma$ efficiency, bolometric correction factor, and the pion suppression factors respectively. For the relevant timescales $f_{p\gamma} \approx f_{\rm bol} \approx f_{\pi,\rm sup} \approx 1$, giving $E_\nu^2 J_\nu \sim 0.003\ {\rm GeV cm}^{-2}$. The $pp$ component in the ejecta also peaks in the energy range $10^7 - 10^8$ GeV but at a fluence $\sim 3 \times 10^{-4}\ {\rm GeV cm}^{-2}$, an order of magnitude lower than the $p\gamma$ component in the nebula. However at late times, $t \gtrsim 10^{6.5}$ s the $pp$ component dominates the neutrino fluence for $E_\nu \gtrsim 10^8$ GeV while for $t\gtrsim 10^7$ s, the $pp$ component dominates over the $p\gamma$ component across almost the entire energy range.

For a single source, that is, a canonical SLSNe at 100 Mpc, the detection prospects with upcoming neutrino telescopes like IceCube-Gen2 or GRAND-200k appear dim. However, this problem can be mitigated by combining data from EM observations to perform the neutrino searches. The current era of powerful EM telescopes like Rubin LSST is ideal for searches of optical transients, like SLSNe. This allows for construction of data catalogs in various individual or a combination of EM wavelengths. These catalogs can then be used to perform maximum log-likelihood searches, where the likelihood functions are designed based on spatial, temporal, and energy correlations between the neutrino and the EM datasets, effectively \emph{stacking} neutrino observations across multiple sources (see for example~\citealt{Mukhopadhyay:2023niv,Mukhopadhyay:2024lwq,Mukhopadhyay:2026qlq}).

A simple estimate of the results for stacking of neutrino observations using a population of observed SLSNe in Rubin LSST within redshift $z_{\rm max} = 1$ is presented in the \emph{right} panel of Figure~\ref{fig:nu}. We show the detection significance $\sigma_{\rm det}$, for a given neutrino telescope as a function of the observation duration $T_{\rm obs}$, where we assume given the field of view of Rubin LSST, it can observe a fraction $1/3$ of all SLSNe at $z_{\rm max} \le 1$. Details regarding the computation of $\sigma_{\rm det}$ are provided in Appendix~\ref{appsec:nu_stack}. We note that the proposed $\sim 10\ {\rm km}^3$ neutrino telescopes like HUNT or TRIDENT can reach $3\sigma$ ($5\sigma$) detection significance in less than $5$ ($20$) years, while upcoming neutrino detectors like IceCube-Gen2 with the optical and radio arrays can detect several signal neutrino events from a population of SLSNe within $10$ years of observation and reach $3\sigma$ detection significance in $\sim 20$ years independently. Since the observed neutrino fluence peaks between $10^7 - 10^8$ GeV, detectors like IceCube-Gen2 and HUNT/TRIDENT with an effective area maximized across the relevant energies provides the best observational scenario.

It is also important to carefully choose the maximum redshift of the population one should consider for stacking searches. The number of SLSNe obviously increases with $z$, but beyond a critical redshift $z_{\rm max} \sim 1$ the additional signal from more distant sources is not sufficient to overcome the background. To avoid a resulting reduction in the detection significance, this necessitates that optical observations of SLSNe also include redshift information from spectroscopic observations or photometric host galaxy information to filter the sample used in stacking analysis for neutrinos. Rubin LSST will provide photometric data which would need to be complemented by current generation telescopes like Subaru~\citep{Tamura:2016wsg}, Gemini, Keck or next-generation ones like European Extremely Large Telescope (E-ELT)~\citep{evans2015science}, US Extremely Large Telescope (US-ELT)~\citep{Graham:2019mis,Milisavljevic:2019jld} comprising of Thirty Meter Telescope (TMT) and the Giant Magellan Telescope (GMT) to obtain the spectroscopic observations.
\section{Conclusions and Outlook}
\label{sec:disc}
We address two main themes in this work -- (a) a prediction for the the multi-wavelength EM and neutrino signatures from a canonical magnetar-powered SLSN and (b) highlighting the complementarity of the EM observations given their excellent detection prospects across wavelengths (particularly optical/IR band at Rubin and Roman) to search for neutrino signatures from a population of SLSNe. 

Among the possible energy sources behind SLSNe is a millisecond magnetar, that is a remnant neutron star with a millisecond spin period and strong dipole magnetic field $10^{13} - 10^{15}$ G. We modeled our system by considering observational data from the nearest observed SLSNe SN 2017egm, in particular the bolometric luminosity and thermal temperature (see Figure~\ref{fig:obs_sn2017egm}). The properties of the ejecta, including mass and composition, were chosen based on numerical simulations of ejecta resulting from core-collapse. Remarkably, our model's predictions for high-energy (GeV) gamma-rays are consistent with recent observational data from Fermi LAT (see Figure~\ref{fig:obs_sn2017egm} \emph{lower panel}), strengthening the conclusions of \citet{Li:2024ics}.

The magnetar's rotational energy serves as the main reservoir for powering all thermal, non-thermal, and neutrino emissions (see Figure~\ref{fig:energies}). This is achieved by systematically solving for the dynamics of the magnetar-wind-nebula-ejecta system and then computing the resulting leptonic and hadronic cascades and interactions. In particular, we also improve several aspects of our model at the microphysics level. Instead of arbitrarily fixing the break of the pair injection spectra, we consistently evaluate the bulk Lorentz factor of the wind to inject the non-thermal pairs accordingly. Furthermore, we also compute the pair multiplicity in the wind and nebula region rather than choosing a fixed value. These results are highlighted in Appendices~\ref{appsubsec:pairs} and~\ref{appsubsec:wind} and provide more realistic observational signatures.

Even with the improvements to our model included here, it still makes several simplifying approximations. Firstly our model computes the lepto-hadronic interactions in an one-zone approximation. A more detailed computation can be performed by considering multi-zone emission models, where the photosphere evolves with time, which can be achieved by considering a spatially varying density profile ($\rho_{\rm ej} \propto r^{-\delta}$) and thermal energy density ($\varepsilon_{\rm th} \propto r^{-\Gamma}$) associated with the ejecta. This would presumably enable a better fit to the photosphere temperature $T_{\rm th}$ at the cost of introducing additional parameters for the ejecta density structure. The ionization state of the ejecta is another uncertainty associated with our modeling. For simplicity we consider singly-ionized ejecta, but recombination of the ejecta as it expands and cools could lead to more optimistic radio signatures. Radio signatures can also be boosted as a result of emission from older pairs in the nebula \citep{Margalit&Metzger18} or from shock breakout~\citep{Chen:2026mnf}. We leave these for future work.

We present the results of theme (a) in Figures~\ref{fig:emspectra} and~\ref{fig:lc} for the EM signatures and Figure~\ref{fig:nu} \emph{(left panel)} for the neutrino signatures. The results corresponding to theme (b) are shown in the \emph{right} panel of Figure~\ref{fig:nu}. We show that stacking searches using a population of SLSNe collected from future EM surveys such as LSST, enable proposed and upcoming neutrino telescopes to detect several signal events over a timescale of $\sim 10$ years; in particular, HUNT/TRIDENT can reach $3\sigma$ and $5\sigma$ detection significance in less than $5$ and $20$ years respectively. IceCube-Gen2 Radio and GRAND can also independently reach $3\sigma$ detection significance in roughly 2 decades of observation time. The detection of neutrinos from a population of SLSNe will take us a step forward in understanding the dynamics and most importantly in unraveling the energy source for SLSNe in the very near future. The current work brings forth the possibilities to gain deeper insights into the engines of SLSNe, enabled by the joint progress in optical transient and neutrino astronomy.
\begin{acknowledgements}
We thank Kazumi Kashiyama for useful discussions. M.\,M acknowledges support from the FermiForward Discovery Group, LLC under Contract No. 89243024CSC000002 with the U.S. Department of Energy, Office of Science, Office of High Energy Physics. S.\,S.\,K acknowledges the support by KAKENHI No. 22K14028, No. 21H04487, No. 23H04899, and the Tohoku Initiative for Fostering Global Researchers for Interdisciplinary Sciences (TI-FRIS) of MEXT’s Strategic Professional Development Program for Young Researchers.  B.\,D.\,M acknowledges support from the National Science Foundation (grant AST-2406637), NASA (grant 80NSSC26K0299), and the Simons Foundation (grant 727700). The Flatiron Institute is supported by the Simons Foundation. I\,.V acknowledges support by the Estonian Research Council (grant PRG2159), Estonian Ministry of Education and Research (grant TK202), and the European Union's Horizon Europe research and innovation programme (EXCOSM, grant No. 101159513).
\end{acknowledgements}
\bibliography{refs}{}

\begin{thebibliography}{}
\expandafter\ifx\csname natexlab\endcsname\relax\def\natexlab#1{#1}\fi
\providecommand{\url}[1]{\href{#1}{#1}}
\providecommand{\dodoi}[1]{doi:~\href{http://doi.org/#1}{\nolinkurl{#1}}}
\providecommand{\doeprint}[1]{\href{http://ascl.net/#1}{\nolinkurl{http://ascl.net/#1}}}
\providecommand{\doarXiv}[1]{\href{https://arxiv.org/abs/#1}{\nolinkurl{https://arxiv.org/abs/#1}}}

\bibitem[{M.~G. Aartsen {et~al.}(2021)Aartsen {et~al.}}]{IceCube-Gen2:2020qha}
Aartsen, M.~G., {et~al.} 2021, \bibinfo{title}{{IceCube-Gen2: the window to the extreme Universe},} J. Phys. G, 48, 060501, \dodoi{10.1088/1361-6471/abbd48}

\bibitem[{R. Abbasi {et~al.}(2021)Abbasi {et~al.}}]{IceCube:2020wum}
Abbasi, R., {et~al.} 2021, \bibinfo{title}{{The IceCube high-energy starting event sample: Description and flux characterization with 7.5 years of data},} Phys. Rev. D, 104, 022002, \dodoi{10.1103/PhysRevD.104.022002}

\bibitem[{R. Abbasi {et~al.}(2023)Abbasi {et~al.}}]{IceCube:2023esf}
Abbasi, R., {et~al.} 2023, \bibinfo{title}{{Constraining High-energy Neutrino Emission from Supernovae with IceCube},} Astrophys. J. Lett., 949, L12, \dodoi{10.3847/2041-8213/acd2c9}

\bibitem[{R. Abbasi {et~al.}(2024)Abbasi {et~al.}}]{IceCube:2024fxo}
Abbasi, R., {et~al.} 2024, \bibinfo{title}{{Characterization of the astrophysical diffuse neutrino flux using starting track events in IceCube},} Phys. Rev. D, 110, 022001, \dodoi{10.1103/PhysRevD.110.022001}

\bibitem[{F. Acero {et~al.}(2026)Acero {et~al.}}]{Fermi-LAT:2026ots}
Acero, F., {et~al.} 2026, \bibinfo{title}{{Gamma-ray signature of superluminous supernovae: Fermi-LAT GeV detection of SN 2017egm and evidence of a central engine},} Astron. Astrophys., 709, A229, \dodoi{10.1051/0004-6361/202558547}

\bibitem[{S. Adrian-Martinez {et~al.}(2016)Adrian-Martinez {et~al.}}]{KM3Net:2016zxf}
Adrian-Martinez, S., {et~al.} 2016, \bibinfo{title}{{Letter of intent for KM3NeT 2.0},} J. Phys. G, 43, 084001, \dodoi{10.1088/0954-3899/43/8/084001}

\bibitem[{J. {\'A}lvarez-Mu{\~n}iz {et~al.}(2020){\'A}lvarez-Mu{\~n}iz {et~al.}}]{GRAND:2018iaj}
{\'A}lvarez-Mu{\~n}iz, J., {et~al.} 2020, \bibinfo{title}{{The Giant Radio Array for Neutrino Detection (GRAND): Science and Design},} Sci. China Phys. Mech. Astron., 63, 219501, \dodoi{10.1007/s11433-018-9385-7}

\bibitem[{J. Arons(2003)Arons}]{Arons:2002yj}
Arons, J. 2003, \bibinfo{title}{{Magnetars in the metagalaxy: an origin for ultrahigh-energy cosmic rays in the nearby universe},} Astrophys. J., 589, 871, \dodoi{10.1086/374776}

\bibitem[{B. Blum {et~al.}(2022)Blum {et~al.}}]{Blum:2022dxi}
Blum, B., {et~al.} 2022, \bibinfo{title}{{Snowmass2021 Cosmic Frontier White Paper: Rubin Observatory after LSST},} in {Snowmass 2021}.
\newblock \doarXiv{2203.07220}

\bibitem[{S. Bose {et~al.}(2018)Bose {et~al.}}]{Bose:2017cgk}
Bose, S., {et~al.} 2018, \bibinfo{title}{{Gaia17biu/SN 2017egm in NGC 3191: The Closest Hydrogen-poor Superluminous Supernova to Date Is in a {\textquotedblleft}Normal,{\textquotedblright} Massive, Metal-rich Spiral Galaxy},} Astrophys. J., 853, 57, \dodoi{10.3847/1538-4357/aaa298}

\bibitem[{D.~N. Burrows {et~al.}(2005)Burrows {et~al.}}]{SWIFT:2005ngz}
Burrows, D.~N., {et~al.} 2005, \bibinfo{title}{{The Swift X-ray Telescope},} Space Sci. Rev., 120, 165, \dodoi{10.1007/s11214-005-5097-2}

\bibitem[{M. Chen(2026)Chen}]{Chen:2026jdx}
Chen, M. 2026, \bibinfo{title}{{HUNT: An ultra-large-scale neutrino astronomy telescope},} Nucl. Instrum. Meth. A, 1086, 171374, \dodoi{10.1016/j.nima.2026.171374}

\bibitem[{M. Chen {et~al.}(2026)Chen, Kashiyama, \& Sato}]{Chen:2026mnf}
Chen, M., Kashiyama, K., \& Sato, M. 2026, \bibinfo{title}{{Blowouts of Nascent Wind Bubbles in Pulsar-Driven Supernovae},} \doarXiv{2601.09552}

\bibitem[{Z.~H. {Chen} {et~al.}(2023){Chen}, {Yan}, {Kangas}, {Lunnan}, {Schulze}, {Sollerman}, {Perley}, {Chen}, {Taggart}, {Hinds}, {Gal-Yam}, {Wang}, {Andreoni}, {Bellm}, {Bloom}, {Burdge}, {Burgos}, {Cook}, {Dahiwale}, {De}, {Dekany}, {Dugas}, {Frederik}, {Fremling}, {Graham}, {Hankins}, {Ho}, {Jencson}, {Karambelkar}, {Kasliwal}, {Kulkarni}, {Laher}, {Rusholme}, {Sharma}, {Taddia}, {Tartaglia}, {Thomas}, {Tzanidakis}, {Van Roestel}, {Walter}, {Yang}, {Yao}, \& {Yaron}}]{Chen+23}
{Chen}, Z.~H., {Yan}, L., {Kangas}, T., {et~al.} 2023, \bibinfo{title}{{The Hydrogen-poor Superluminous Supernovae from the Zwicky Transient Facility Phase I Survey. I. Light Curves and Measurements},} \apj, 943, 41, \dodoi{10.3847/1538-4357/aca161}

\bibitem[{R.~A. {Chevalier} \& C.~M. {Irwin}(2011){Chevalier} \& {Irwin}}]{2011ApJ...729L...6C}
{Chevalier}, R.~A., \& {Irwin}, C.~M. 2011, \bibinfo{title}{{Shock Breakout in Dense Mass Loss: Luminous Supernovae},} \apjl, 729, L6, \dodoi{10.1088/2041-8205/729/1/L6}

\bibitem[{M. Crnogor{\v{c}}evi{\'c} {et~al.}(2026)Crnogor{\v{c}}evi{\'c}, Linden, Goobar, \& Metzger}]{Crnogorcevic:2026alv}
Crnogor{\v{c}}evi{\'c}, M., Linden, T., Goobar, A., \& Metzger, B.~D. 2026, \bibinfo{title}{{On the Gamma-ray Efficiency of Superluminous Supernovae: Potential Detections and Population-Level Constraints},} \doarXiv{2604.16595}

\bibitem[{A. {Delgado} {et~al.}(2017){Delgado}, {Harrison}, {Hodgkin}, {Leeuwen}, {Rixon}, \& {Yoldas}}]{2017TNSTR.591....1D}
{Delgado}, A., {Harrison}, D., {Hodgkin}, S., {et~al.} 2017, \bibinfo{title}{{GaiaAlerts Transient Discovery Report for 2017-05-25},} Transient Name Server Discovery Report, 2017-591, 1

\bibitem[{L. Dessart {et~al.}(2012)Dessart, Hillier, Waldman, Livne, \& Blondin}]{Dessart:2012vc}
Dessart, L., Hillier, D.~J., Waldman, R., Livne, E., \& Blondin, S. 2012, \bibinfo{title}{{Super-luminous supernovae: 56Ni power versus magnetar radiation},} Mon. Not. Roy. Astron. Soc., 426, L76, \dodoi{10.1111/j.1745-3933.2012.01329.x}

\bibitem[{L. Dessart {et~al.}(2015)Dessart, Hillier, Woosley, Livne, Waldman, Yoon, \& Langer}]{Dessart:2015mga}
Dessart, L., Hillier, D.~J., Woosley, S., {et~al.} 2015, \bibinfo{title}{{Radiative-transfer models for supernovae IIb/Ib/Ic from binary-star progenitors},} Mon. Not. Roy. Astron. Soc., 453, 2189, \dodoi{10.1093/mnras/stv1747}

\bibitem[{L. Dessart {et~al.}(2016)Dessart, Hillier, Woosley, Livne, Waldman, Yoon, \& Langer}]{Dessart:2016fun}
Dessart, L., Hillier, D.~J., Woosley, S., {et~al.} 2016, \bibinfo{title}{{Inferring supernova IIb/Ib/Ic ejecta properties from light curves and spectra: Correlations from radiative-transfer models},} Mon. Not. Roy. Astron. Soc., 458, 1618, \dodoi{10.1093/mnras/stw418}

\bibitem[{T. Eftekhari {et~al.}(2021)Eftekhari {et~al.}}]{Eftekhari:2020ynh}
Eftekhari, T., {et~al.} 2021, \bibinfo{title}{{Late-time Radio and Millimeter Observations of Superluminous Supernovae and Long Gamma-Ray Bursts: Implications for Central Engines, Fast Radio Bursts, and Obscured Star Formation},} Astrophys. J., 912, 21, \dodoi{10.3847/1538-4357/abe9b8}

\bibitem[{N. Ekanger {et~al.}(2026)Ekanger, Kimura, \& Kashiyama}]{Ekanger:2026coc}
Ekanger, N., Kimura, S.~S., \& Kashiyama, K. 2026, \bibinfo{title}{{Super-knee cosmic rays from interacting supernovae},} \doarXiv{2602.06410}

\bibitem[{R. Enberg {et~al.}(2008)Enberg, Reno, \& Sarcevic}]{Enberg:2008te}
Enberg, R., Reno, M.~H., \& Sarcevic, I. 2008, \bibinfo{title}{{Prompt neutrino fluxes from atmospheric charm},} Phys. Rev. D, 78, 043005, \dodoi{10.1103/PhysRevD.78.043005}

\bibitem[{C. Evans {et~al.}(2015)Evans, Puech, Afonso, Almaini, Amram, Aussel, Barbuy, Basden, Bastian, Battaglia, {et~al.}}]{evans2015science}
Evans, C., Puech, M., Afonso, J., {et~al.} 2015, \bibinfo{title}{The science case for multi-object spectroscopy on the European ELT,} arXiv preprint arXiv:1501.04726

\bibitem[{K. {Fang} \& B.~D. {Metzger}(2017){Fang} \& {Metzger}}]{Fang&Metzger17}
{Fang}, K., \& {Metzger}, B.~D. 2017, \bibinfo{title}{{High-energy Neutrinos from Millisecond Magnetars Formed from the Merger of Binary Neutron Stars},} \apj, 849, 153, \dodoi{10.3847/1538-4357/aa8b6a}

\bibitem[{K. {Fang} {et~al.}(2019){Fang}, {Metzger}, {Murase}, {Bartos}, \& {Kotera}}]{Fang+19}
{Fang}, K., {Metzger}, B.~D., {Murase}, K., {Bartos}, I., \& {Kotera}, K. 2019, \bibinfo{title}{{Multimessenger Implications of AT2018cow: High-energy Cosmic-Ray and Neutrino Emissions from Magnetar-powered Superluminous Transients},} \apj, 878, 34, \dodoi{10.3847/1538-4357/ab1b72}

\bibitem[{A. Gal-Yam(2019)Gal-Yam}]{Gal-Yam:2018out}
Gal-Yam, A. 2019, \bibinfo{title}{{The Most Luminous Supernovae},} Ann. Rev. Astron. Astrophys., 57, 305, \dodoi{10.1146/annurev-astro-081817-051819}

\bibitem[{A. {Gal-Yam} \& D.~C. {Leonard}(2009){Gal-Yam} \& {Leonard}}]{2009Natur.458..865G}
{Gal-Yam}, A., \& {Leonard}, D.~C. 2009, \bibinfo{title}{{A massive hypergiant star as the progenitor of the supernova SN 2005gl},} \nat, 458, 865, \dodoi{10.1038/nature07934}

\bibitem[{A. {Gal-Yam} {et~al.}(2009){Gal-Yam}, {Mazzali}, {Ofek}, {Nugent}, {Kulkarni}, {Kasliwal}, {Quimby}, {Filippenko}, {Cenko}, {Chornock}, {Waldman}, {Kasen}, {Sullivan}, {Beshore}, {Drake}, {Thomas}, {Bloom}, {Poznanski}, {Miller}, {Foley}, {Silverman}, {Arcavi}, {Ellis}, \& {Deng}}]{2009Natur.462..624G}
{Gal-Yam}, A., {Mazzali}, P., {Ofek}, E.~O., {et~al.} 2009, \bibinfo{title}{{Supernova 2007bi as a pair-instability explosion},} \nat, 462, 624, \dodoi{10.1038/nature08579}

\bibitem[{P. Goldreich \& W.~H. Julian(1969)Goldreich \& Julian}]{Goldreich:1969sb}
Goldreich, P., \& Julian, W.~H. 1969, \bibinfo{title}{{Pulsar electrodynamics},} Astrophys. J., 157, 869, \dodoi{10.1086/150119}

\bibitem[{S. {Gomez} {et~al.}(2024){Gomez}, {Nicholl}, {Berger}, {Blanchard}, {Villar}, {Rest}, {Hosseinzadeh}, {Aamer}, {Ajay}, {Athukoralalage}, {Coulter}, {Eftekhari}, {Fiore}, {Franz}, {Fox}, {Gagliano}, {Hiramatsu}, {Howell}, {Hsu}, {Karmen}, {Siebert}, {K{\"o}nyves-T{\'o}th}, {Kumar}, {McCully}, {Pellegrino}, {Pierel}, {Rest}, \& {Wang}}]{Gomez+24}
{Gomez}, S., {Nicholl}, M., {Berger}, E., {et~al.} 2024, \bibinfo{title}{{The Type I superluminous supernova catalogue I: light-curve properties, models, and catalogue description},} \mnras, 535, 471, \dodoi{10.1093/mnras/stae2270}

\bibitem[{M.~L. Graham {et~al.}(2019)Graham {et~al.}}]{Graham:2019mis}
Graham, M.~L., {et~al.} 2019, \bibinfo{title}{{Astro2020 Science White Paper: Discovery Frontiers of Explosive Transients - An ELT {\textbackslash}{\&} LSST Perspective},} \doarXiv{1904.05957}

\bibitem[{D. Harari {et~al.}(2014)Harari, Mollerach, \& Roulet}]{Harari:2013pea}
Harari, D., Mollerach, S., \& Roulet, E. 2014, \bibinfo{title}{{Anisotropies of ultrahigh energy cosmic rays diffusing from extragalactic sources},} Phys. Rev. D, 89, 123001, \dodoi{10.1103/PhysRevD.89.123001}

\bibitem[{M. Honda {et~al.}(2007)Honda, Kajita, Kasahara, Midorikawa, \& Sanuki}]{Honda:2006qj}
Honda, M., Kajita, T., Kasahara, K., Midorikawa, S., \& Sanuki, T. 2007, \bibinfo{title}{{Calculation of atmospheric neutrino flux using the interaction model calibrated with atmospheric muon data},} Phys. Rev. D, 75, 043006, \dodoi{10.1103/PhysRevD.75.043006}

\bibitem[{ {IceCube-Gen2 Collaboration}(2024){IceCube-Gen2 Collaboration}}]{Gen2_TDR}
{IceCube-Gen2 Collaboration}. 2024, {IceCube-Gen2 Technical Design Report}, Tech. rep.
\newblock \url{https://icecube-gen2.wisc.edu/science/publications/tdr/}

\bibitem[{{\v{Z}}. Ivezi{\'c} {et~al.}(2019)Ivezi{\'c} {et~al.}}]{LSST:2008ijt}
Ivezi{\'c}, {\v{Z}}., {et~al.} 2019, \bibinfo{title}{{LSST: from Science Drivers to Reference Design and Anticipated Data Products},} Astrophys. J., 873, 111, \dodoi{10.3847/1538-4357/ab042c}

\bibitem[{D. Kasen \& L. Bildsten(2010)Kasen \& Bildsten}]{Kasen:2009tg}
Kasen, D., \& Bildsten, L. 2010, \bibinfo{title}{{Supernova Light Curves Powered by Young Magnetars},} Astrophys. J., 717, 245, \dodoi{10.1088/0004-637X/717/1/245}

\bibitem[{B. Katz {et~al.}(2012)Katz, Sapir, \& Waxman}]{Katz:2011fz}
Katz, B., Sapir, N., \& Waxman, E. 2012, \bibinfo{title}{{Non-relativistic radiation mediated shock breakouts: II. Bolometric properties of SN shock breakout},} Astrophys. J., 747, 147, \dodoi{10.1088/0004-637X/747/2/147}

\bibitem[{S.~R. Kelner \& F.~A. Aharonian(2008)Kelner \& Aharonian}]{Kelner:2008ke}
Kelner, S.~R., \& Aharonian, F.~A. 2008, \bibinfo{title}{{Energy spectra of gamma-rays, electrons and neutrinos produced at interactions of relativistic protons with low energy radiation},} Phys. Rev. D, 78, 034013, \dodoi{10.1103/PhysRevD.82.099901}

\bibitem[{S.~R. Kelner {et~al.}(2006)Kelner, Aharonian, \& Bugayov}]{Kelner:2006tc}
Kelner, S.~R., Aharonian, F.~A., \& Bugayov, V.~V. 2006, \bibinfo{title}{{Energy spectra of gamma-rays, electrons and neutrinos produced at proton-proton interactions in the very high energy regime},} Phys. Rev. D, 74, 034018, \dodoi{10.1103/PhysRevD.74.034018}

\bibitem[{A. Kheirandish \& K. Murase(2023)Kheirandish \& Murase}]{Kheirandish:2022eox}
Kheirandish, A., \& Murase, K. 2023, \bibinfo{title}{{Detecting High-energy Neutrino Minibursts from Local Supernovae with Multiple Neutrino Observatories},} Astrophys. J. Lett., 956, L8, \dodoi{10.3847/2041-8213/acf84f}

\bibitem[{S.~S. Kimura {et~al.}(2019)Kimura, Murase, \& M{\'e}sz{\'a}ros}]{Kimura:2019yjo}
Kimura, S.~S., Murase, K., \& M{\'e}sz{\'a}ros, P. 2019, \bibinfo{title}{{Multimessenger tests of cosmic-ray acceleration in radiatively inefficient accretion flows},} Phys. Rev. D, 100, 083014, \dodoi{10.1103/PhysRevD.100.083014}

\bibitem[{K. Kotera {et~al.}(2026)Kotera, Mukhopadhyay, Alves~Batista, Fox, Martineau-Huynh, Murase, Wissel, \& Zeolla}]{Kotera:2025jca}
Kotera, K., Mukhopadhyay, M., Alves~Batista, R., {et~al.} 2026, \bibinfo{title}{{Observational strategies for ultrahigh-energy neutrinos: the importance of deep sensitivity for detection and astronomy},} JCAP, 01, 027, \dodoi{10.1088/1475-7516/2026/01/027}

\bibitem[{K. Kotera {et~al.}(2013)Kotera, Phinney, \& Olinto}]{Kotera:2013yaa}
Kotera, K., Phinney, E.~S., \& Olinto, A.~V. 2013, \bibinfo{title}{{Signatures of pulsars in the light curves of newly formed supernova remnants},} Mon. Not. Roy. Astron. Soc., 432, 3228, \dodoi{10.1093/mnras/stt680}

\bibitem[{S. Li {et~al.}(2024)Li, Liang, Liao, Lei, \& Fan}]{Li:2024ics}
Li, S., Liang, Y.-F., Liao, N.-H., Lei, L., \& Fan, Y.-Z. 2024, \bibinfo{title}{{Evidence for GeV emission of the superluminous supernova SN 2017egm},} \doarXiv{2407.05968}

\bibitem[{B. {Margalit} \& B.~D. {Metzger}(2018){Margalit} \& {Metzger}}]{Margalit&Metzger18}
{Margalit}, B., \& {Metzger}, B.~D. 2018, \bibinfo{title}{{A Concordance Picture of FRB 121102 as a Flaring Magnetar Embedded in a Magnetized Ion-Electron Wind Nebula},} \apjl, 868, L4, \dodoi{10.3847/2041-8213/aaedad}

\bibitem[{B. Margalit {et~al.}(2018)Margalit, Metzger, Berger, Nicholl, Eftekhari, \& Margutti}]{Margalit:2018bje}
Margalit, B., Metzger, B.~D., Berger, E., {et~al.} 2018, \bibinfo{title}{{Unveiling the engines of fast radio bursts, superluminous supernovae, and gamma-ray bursts},} Mon. Not. Roy. Astron. Soc., 481, 2407, \dodoi{10.1093/mnras/sty2417}

\bibitem[{B.~D. Metzger {et~al.}(2017)Metzger, Berger, \& Margalit}]{Metzger:2017wdz}
Metzger, B.~D., Berger, E., \& Margalit, B. 2017, \bibinfo{title}{{Millisecond Magnetar Birth Connects FRB 121102 to Superluminous Supernovae and Long Duration Gamma-ray Bursts},} Astrophys. J., 841, 14, \dodoi{10.3847/1538-4357/aa633d}

\bibitem[{B.~D. {Metzger} \& A.~L. {Piro}(2014){Metzger} \& {Piro}}]{Metzger&Piro14}
{Metzger}, B.~D., \& {Piro}, A.~L. 2014, \bibinfo{title}{{Optical and X-ray emission from stable millisecond magnetars formed from the merger of binary neutron stars},} \mnras, 439, 3916, \dodoi{10.1093/mnras/stu247}

\bibitem[{B.~D. Metzger {et~al.}(2014)Metzger, Vurm, Hascoet, \& Beloborodov}]{Metzger:2013kia}
Metzger, B.~D., Vurm, I., Hascoet, R., \& Beloborodov, A.~M. 2014, \bibinfo{title}{{Ionization Break-Out from Millisecond Pulsar Wind Nebulae: an X-ray Probe of the Origin of Superluminous Supernovae},} Mon. Not. Roy. Astron. Soc., 437, 703, \dodoi{10.1093/mnras/stt1922}

\bibitem[{D. Milisavljevic {et~al.}(2019)Milisavljevic {et~al.}}]{Milisavljevic:2019jld}
Milisavljevic, D., {et~al.} 2019, \bibinfo{title}{{Achieving Transformative Understanding of Extreme Stellar Explosions with ELT-enabled Late-time Spectroscopy},} \doarXiv{1904.05897}

\bibitem[{T.~J. Moriya(2024)Moriya}]{Moriya:2024gqt}
Moriya, T.~J. 2024, \bibinfo{title}{{Superluminous supernovae},} \doarXiv{2407.12302}

\bibitem[{M. Mukhopadhyay \& S.~S. Kimura(2025)Mukhopadhyay \& Kimura}]{Mukhopadhyay:2025tvz}
Mukhopadhyay, M., \& Kimura, S.~S. 2025, \bibinfo{title}{{Electromagnetic Signatures from Pulsar Remnants of Binary Neutron Star Mergers: Prospects for Unique Identification Using Multiwavelength Signatures},} Astrophys. J. Lett., 989, L41, \dodoi{10.3847/2041-8213/adf285}

\bibitem[{M. Mukhopadhyay \& S.~S. Kimura(2026)Mukhopadhyay \& Kimura}]{Mukhopadhyay:2026wrv}
Mukhopadhyay, M., \& Kimura, S.~S. 2026, \bibinfo{title}{{High energy neutrinos from pulsar-powered optical transients: LFBOTs as potential origin of the KM3NeT event KM3-230213A},} \doarXiv{2601.22266}

\bibitem[{M. Mukhopadhyay {et~al.}(2025)Mukhopadhyay, Kimura, \& Metzger}]{Mukhopadhyay:2024ehs}
Mukhopadhyay, M., Kimura, S.~S., \& Metzger, B.~D. 2025, \bibinfo{title}{{High-energy neutrino signatures from pulsar remnants of binary neutron-star mergers: coincident detection prospects with gravitational waves},} Astrophys. J., 987, 2, \dodoi{10.3847/1538-4357/adc913}

\bibitem[{M. Mukhopadhyay {et~al.}(2024{\natexlab{a}})Mukhopadhyay, Kimura, \& Murase}]{Mukhopadhyay:2023niv}
Mukhopadhyay, M., Kimura, S.~S., \& Murase, K. 2024{\natexlab{a}}, \bibinfo{title}{{Gravitational wave triggered searches for high-energy neutrinos from binary neutron star mergers: Prospects for next generation detectors},} Phys. Rev. D, 109, 043053, \dodoi{10.1103/PhysRevD.109.043053}

\bibitem[{M. Mukhopadhyay {et~al.}(2024{\natexlab{b}})Mukhopadhyay, Kotera, Wissel, Murase, \& Kimura}]{Mukhopadhyay:2024lwq}
Mukhopadhyay, M., Kotera, K., Wissel, S., Murase, K., \& Kimura, S.~S. 2024{\natexlab{b}}, \bibinfo{title}{{Ultrahigh-energy neutrino searches using next-generation gravitational wave detectors at radio neutrino detectors: GRAND, IceCube-Gen2 Radio, and RNO-G},} Phys. Rev. D, 110, 063004, \dodoi{10.1103/PhysRevD.110.063004}

\bibitem[{M. Mukhopadhyay {et~al.}(2026)Mukhopadhyay, Wusinich, \& Murase}]{Mukhopadhyay:2026qlq}
Mukhopadhyay, M., Wusinich, P., \& Murase, K. 2026, \bibinfo{title}{{Constraining high-energy neutrinos from tidal disruption events with IceCube high-energy starting events},} \doarXiv{2601.20934}

\bibitem[{K. Murase(2024)Murase}]{Murase:2023chr}
Murase, K. 2024, \bibinfo{title}{{Interacting supernovae as high-energy multimessenger transients},} Phys. Rev. D, 109, 103020, \dodoi{10.1103/PhysRevD.109.103020}

\bibitem[{K. Murase {et~al.}(2015)Murase, Kashiyama, Kiuchi, \& Bartos}]{Murase:2014bfa}
Murase, K., Kashiyama, K., Kiuchi, K., \& Bartos, I. 2015, \bibinfo{title}{{Gamma-Ray and Hard X-Ray Emission from Pulsar-Aided Supernovae as a Probe of Particle Acceleration in Embryonic Pulsar Wind Nebulae},} Astrophys. J., 805, 82, \dodoi{10.1088/0004-637X/805/1/82}

\bibitem[{K. Murase {et~al.}(2016)Murase, Kashiyama, \& M{\'e}sz{\'a}ros}]{Murase:2016sqo}
Murase, K., Kashiyama, K., \& M{\'e}sz{\'a}ros, P. 2016, \bibinfo{title}{{A Burst in a Wind Bubble and the Impact on Baryonic Ejecta: High-Energy Gamma-Ray Flashes and Afterglows from Fast Radio Bursts and Pulsar-Driven Supernova Remnants},} Mon. Not. Roy. Astron. Soc., 461, 1498, \dodoi{10.1093/mnras/stw1328}

\bibitem[{K. Murase {et~al.}(2011)Murase, Thompson, Lacki, \& Beacom}]{Murase:2010cu}
Murase, K., Thompson, T.~A., Lacki, B.~C., \& Beacom, J.~F. 2011, \bibinfo{title}{{New Class of High-Energy Transients from Crashes of Supernova Ejecta with Massive Circumstellar Material Shells},} Phys. Rev. D, 84, 043003, \dodoi{10.1103/PhysRevD.84.043003}

\bibitem[{K. Murase {et~al.}(2014)Murase, Thompson, \& Ofek}]{Murase:2013kda}
Murase, K., Thompson, T.~A., \& Ofek, E.~O. 2014, \bibinfo{title}{{Probing Cosmic-Ray Ion Acceleration with Radio-Submm and Gamma-Ray Emission from Interaction-Powered Supernovae},} Mon. Not. Roy. Astron. Soc., 440, 2528, \dodoi{10.1093/mnras/stu384}

\bibitem[{K. Murase {et~al.}(2021)Murase {et~al.}}]{Murase:2021lro}
Murase, K., {et~al.} 2021, \bibinfo{title}{{ALMA and NOEMA constraints on synchrotron nebular emission from embryonic superluminous supernova remnants and radio{\textendash}gamma-ray connection},} Mon. Not. Roy. Astron. Soc., 508, 44, \dodoi{10.1093/mnras/stab2506}

\bibitem[{M. Nicholl {et~al.}(2017{\natexlab{a}})Nicholl, Berger, Margutti, Blanchard, Guillochon, Leja, \& Chornock}]{Nicholl:2017mnb}
Nicholl, M., Berger, E., Margutti, R., {et~al.} 2017{\natexlab{a}}, \bibinfo{title}{{The superluminous supernova SN 2017egm in the nearby galaxy NGC 3191: a metal-rich environment can support a typical SLSN evolution},} Astrophys. J. Lett., 845, L8, \dodoi{10.3847/2041-8213/aa82b1}

\bibitem[{M. Nicholl {et~al.}(2017{\natexlab{b}})Nicholl, Guillochon, \& Berger}]{Nicholl_2017}
Nicholl, M., Guillochon, J., \& Berger, E. 2017{\natexlab{b}}, \bibinfo{title}{The Magnetar Model for Type I Superluminous Supernovae. I. Bayesian Analysis of the Full Multicolor Light-curve Sample with MOSFiT,} The Astrophysical Journal, 850, 55, \dodoi{10.3847/1538-4357/aa9334}

\bibitem[{M. Nicholl {et~al.}(2018)Nicholl {et~al.}}]{Nicholl:2018cam}
Nicholl, M., {et~al.} 2018, \bibinfo{title}{{One Thousand Days of SN2015bn: HST Imaging Shows a Light Curve Flattening Consistent with Magnetar Predictions},} Astrophys. J. Lett., 866, L24, \dodoi{10.3847/2041-8213/aae70d}

\bibitem[{E.~O. Ofek {et~al.}(2007)Ofek {et~al.}}]{Ofek:2006vt}
Ofek, E.~O., {et~al.} 2007, \bibinfo{title}{{SN 2006gy: An extremely luminous supernova in the early-type galaxy NGC 1260},} Astrophys. J. Lett., 659, L13, \dodoi{10.1086/516749}

\bibitem[{C.~M.~B. Omand \& N. Sarin(2023)Omand \& Sarin}]{Omand:2023fii}
Omand, C. M.~B., \& Sarin, N. 2023, \bibinfo{title}{{A generalized semi-analytic model for magnetar-driven supernovae},} Mon. Not. Roy. Astron. Soc., 527, 6455, \dodoi{10.1093/mnras/stad3645}

\bibitem[{T. Pitik {et~al.}(2023)Pitik, Tamborra, Lincetto, \& Franckowiak}]{Pitik:2023vcg}
Pitik, T., Tamborra, I., Lincetto, M., \& Franckowiak, A. 2023, \bibinfo{title}{{Optically informed searches of high-energy neutrinos from interaction-powered supernovae},} Mon. Not. Roy. Astron. Soc., 524, 3366, \dodoi{10.1093/mnras/stad2025}

\bibitem[{R.~M. Quimby {et~al.}(2007)Quimby, Aldering, Wheeler, Hoflich, Akerlof, \& Rykoff}]{Quimby:2007tb}
Quimby, R.~M., Aldering, G., Wheeler, J.~C., {et~al.} 2007, \bibinfo{title}{{SN 2005ap: A Most Brilliant Explosion},} Astrophys. J. Lett., 668, L99, \dodoi{10.1086/522862}

\bibitem[{R.~M. Quimby {et~al.}(2011)Quimby {et~al.}}]{Quimby:2009ps}
Quimby, R.~M., {et~al.} 2011, \bibinfo{title}{{Hydrogen-poor superluminous stellar explosions},} Nature, 474, 487, \dodoi{10.1038/nature10095}

\bibitem[{N. Renault-Tinacci {et~al.}(2018)Renault-Tinacci, Kotera, Neronov, \& Ando}]{Renault-Tinacci:2017gon}
Renault-Tinacci, N., Kotera, K., Neronov, A., \& Ando, S. 2018, \bibinfo{title}{{Search for {\ensuremath{\gamma}}-ray emission from superluminous supernovae with the Fermi-LAT},} Astron. Astrophys., 611, A45, \dodoi{10.1051/0004-6361/201730741}

\bibitem[{L. Sironi \& A. Spitkovsky(2011)Sironi \& Spitkovsky}]{Sironi:2011zf}
Sironi, L., \& Spitkovsky, A. 2011, \bibinfo{title}{{Acceleration of Particles at the Termination Shock of a Relativistic Striped Wind},} Astrophys. J., 741, 39, \dodoi{10.1088/0004-637X/741/1/39}

\bibitem[{L. Sironi \& A. Spitkovsky(2014)Sironi \& Spitkovsky}]{Sironi:2014jfa}
Sironi, L., \& Spitkovsky, A. 2014, \bibinfo{title}{{Relativistic Reconnection: an Efficient Source of Non-Thermal Particles},} Astrophys. J. Lett., 783, L21, \dodoi{10.1088/2041-8205/783/1/L21}

\bibitem[{N. Smith {et~al.}(2010)Smith, Chornock, Silverman, Filippenko, \& Foley}]{Smith:2009ce}
Smith, N., Chornock, R., Silverman, J.~M., Filippenko, A.~V., \& Foley, R.~J. 2010, \bibinfo{title}{{Spectral Evolution of the Extraordinary Type IIn Supernova 2006gy},} Astrophys. J., 709, 856, \dodoi{10.1088/0004-637X/709/2/856}

\bibitem[{A. Suzuki \& K. Maeda(2017)Suzuki \& Maeda}]{Suzuki:2016gbg}
Suzuki, A., \& Maeda, K. 2017, \bibinfo{title}{{Supernova ejecta with a relativistic wind from a central compact object: a unified picture for extraordinary supernovae},} Mon. Not. Roy. Astron. Soc., 466, 2633, \dodoi{10.1093/mnras/stw3259}

\bibitem[{R. {Svensson}(1987){Svensson}}]{1987MNRAS.227..403S}
{Svensson}, R. 1987, \bibinfo{title}{{Non-thermal pair production in compact X-ray sources : first-order Compton cascades in soft radiation fields.},} \mnras, 227, 403, \dodoi{10.1093/mnras/227.2.403}

\bibitem[{N. Tamura {et~al.}(2016)Tamura {et~al.}}]{Tamura:2016wsg}
Tamura, N., {et~al.} 2016, \bibinfo{title}{{Prime Focus Spectrograph (PFS) for the Subaru Telescope: Overview, recent progress, and future perspectives},} Proc. SPIE Int. Soc. Opt. Eng., 9908, 99081M, \dodoi{10.1117/12.2232103}

\bibitem[{D.~F. Torres {et~al.}(2014)Torres, Cillis, Mart{\'\i}n, \& de~O{\~n}a~Wilhelmi}]{Torres:2014iua}
Torres, D.~F., Cillis, A., Mart{\'\i}n, J., \& de~O{\~n}a~Wilhelmi, E. 2014, \bibinfo{title}{{Time-dependent modeling of TeV-detected, young pulsar wind nebulae},} JHEAp, 1-2, 31, \dodoi{10.1016/j.jheap.2014.02.001}

\bibitem[{I. Vurm \& B.~D. Metzger(2021)Vurm \& Metzger}]{Vurm:2021dgo}
Vurm, I., \& Metzger, B.~D. 2021, \bibinfo{title}{{Gamma-Ray Thermalization and Leakage from Millisecond Magnetar Nebulae: Toward a Self-consistent Model for Superluminous Supernovae},} Astrophys. J., 917, 77, \dodoi{10.3847/1538-4357/ac0826}

\bibitem[{S.~E. Woosley(2010)Woosley}]{Woosley:2009tu}
Woosley, S.~E. 2010, \bibinfo{title}{{Bright Supernovae from Magnetar Birth},} Astrophys. J. Lett., 719, L204, \dodoi{10.1088/2041-8205/719/2/L204}

\bibitem[{Z.~P. Ye {et~al.}(2023)Ye {et~al.}}]{TRIDENT:2022hql}
Ye, Z.~P., {et~al.} 2023, \bibinfo{title}{{A multi-cubic-kilometre neutrino telescope in the western Pacific Ocean},} Nature Astron., 7, 1497, \dodoi{10.1038/s41550-023-02087-6}

\bibitem[{A.~A. {Zdziarski}(1988){Zdziarski}}]{1988ApJ...335..786Z}
{Zdziarski}, A.~A. 1988, \bibinfo{title}{{Saturated Pair-Photon Cascades on Isotropic Background Photons},} \apj, 335, 786, \dodoi{10.1086/166967}

\bibitem[{H. Zhang {et~al.}(2025)Zhang, Cui, Huang, Lin, Liu, Qiu, Shao, Shi, Xie, \& Yang}]{Zhang:2024slv}
Zhang, H., Cui, Y., Huang, Y., {et~al.} 2025, \bibinfo{title}{{A proposed deep sea Neutrino Observatory in the Nanhai},} Astropart. Phys., 171, 103123, \dodoi{10.1016/j.astropartphys.2025.103123}

\bibitem[{J. Zhu {et~al.}(2023)Zhu {et~al.}}]{Zhu:2023ntt}
Zhu, J., {et~al.} 2023, \bibinfo{title}{{SN 2017egm: A Helium-rich Superluminous Supernova with Multiple Bumps in the Light Curves},} Astrophys. J., 949, 23, \dodoi{10.3847/1538-4357/acc2c3}

\bibitem[{J.-P. Zhu {et~al.}(2024)Zhu, Liu, Yu, Mandel, Hirai, Zhang, \& Chen}]{Zhu:2024yrc}
Zhu, J.-P., Liu, L.-D., Yu, Y.-W., {et~al.} 2024, \bibinfo{title}{{Bumpy Superluminous Supernovae Powered by a Magnetar{\textendash}Star Binary Engine},} Astrophys. J. Lett., 970, L42, \dodoi{10.3847/2041-8213/ad63a8}

\end{thebibliography}
\bibliographystyle{aasjournalv7}
\appendix
\section{Pairs in the wind and nebula}
\label{appsec:wind_neb_pairs}
In this section, we present our calculations for computing the pair multiplicity in the wind and nebular regions and the bulk Lorentz factor in the wind region. The bulk Lorentz factor of the wind (or the effective Lorentz factor for the magnetized scenario) primarily governs the break Lorentz factor of the pair injection. The pair multiplicity helps in estimating the diffusion timescale of the photons in the wind and nebula.
\subsection{Pair Multiplicity in the nebula}
\label{appsubsec:pairs}
\begin{figure}
\centering
\includegraphics[width=0.49\textwidth]{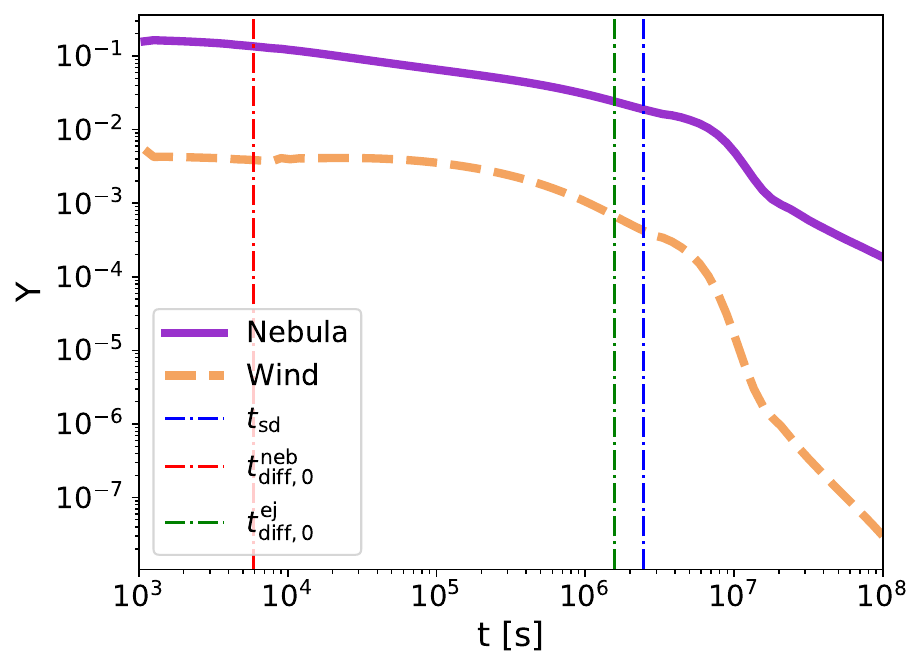}
\includegraphics[width=0.49\textwidth]{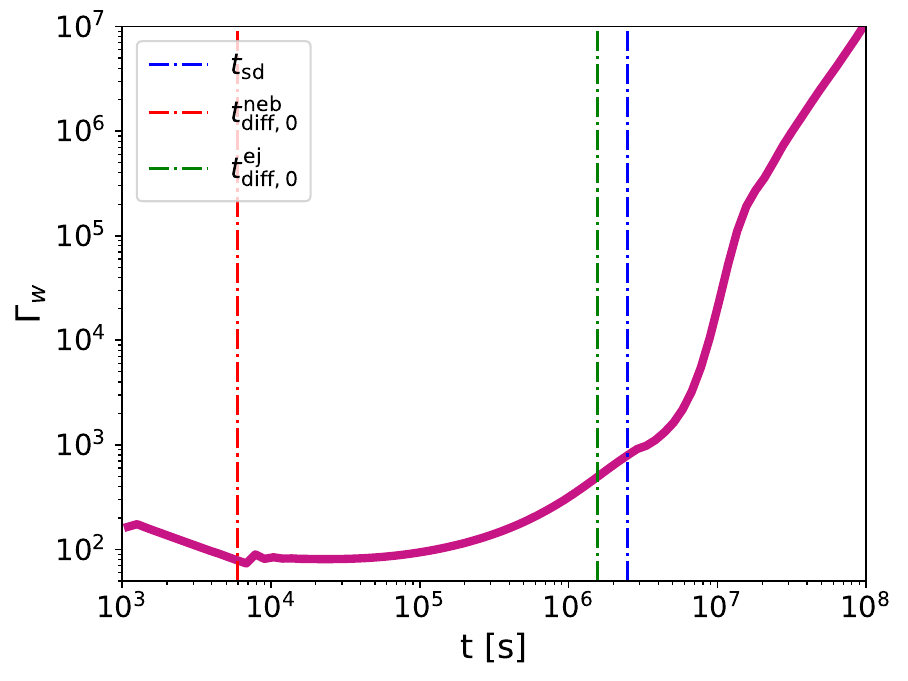}
\caption{\label{fig:pairmult} \emph{Left: }Time evolution of pair multiplicity ($Y$) in the wind and nebula. \emph{Right: }The Lorentz factor of the wind which is also $\gamma_{e,\rm br}$ for the pair injection in the nebula.
}
\end{figure}
We begin by discussing the prescription to evaluate the pair multiplicity. The pair annihilation rate is determined by $\dot{N}_{e^\pm}^{\rm ann} = (3/16)\sigma_T cn_{\pm} N_{\pm}$, where $\sigma_T$ is the Thomson cross-section, $n_\pm$ is the number density of pairs, and the total number of pairs is given by $N_\pm = n_\pm V$, $V$ is the volume of the nebula. The pair annihilation timescale $t_{e^\pm}^{\rm ann}$ can be estimated as, $t_{e^\pm}^{\rm ann} = 1/\big( (3/16)\sigma_T n_{\pm} \big)$. The total number of pairs in the nebula is given by,
\be
\label{eq:pairs_neb}
N_{\pm} = 
\begin{cases}
\left(\frac{\dot{N}_{e^\pm}^{\rm c} V}{(3/16)\sigma_T c}\right)^{1/2}\,, t_{e^\pm}^{\rm ann} \leq t_{\rm dyn}\\
\dot{N}_{e^\pm}^{\rm c} t_{\rm dyn}\,, t_{e^\pm}^{\rm ann} > t_{\rm dyn}
\end{cases}\,,
\ee
where the pair creation rate is given by $\dot{N}_{e^\pm}^{\rm c}$ and $t_{\rm dyn}$ is the dynamical timescale. We compute the pair creation rate by taking into account the pairs created from primary and secondary photons. The pair multiplicity in the nebula is then evaluated as  $Y = \dot{N}_{e^\pm}^{\rm c} m_e c^2/L_{\rm sd}$. The first row of the above equation is obtained by balancing the pair creation and annihilation rates, that is, $\dot{N}_{e^\pm}^{\rm c}=\dot{N}_{e^\pm}^{\rm ann}$ and solving for the total number of pairs. When the pair annihilation rate is longer than the dynamical timescale, the pairs do not have enough time to annihilate.

In Figure~\ref{fig:pairmult} \emph{(left panel)} we show the time evolution of the pair multiplicity. The assumption of $Y = 0.1$ is good for early times but the pair multiplicity drops significantly at late times by an order of magnitude. This happens when $t_{e^\pm}^{\rm ann} > t_{\rm dyn}$.
\subsection{Lorentz factor of the wind $\Gamma_w$}
\label{appsubsec:wind}
In this subsection, we compute the Lorentz factor associated with the wind. The ultra-relativistic wind is slowed down by the ejecta which then thermalizes to form the termination shock (TS). We denote the radial distance of the wind TS from the central engine as $R_s$ and the bulk Lorentz factor of the wind is denoted as $\Gamma_w$.

The rate of energy deposited in the wind can be estimated using $L_w = 4 \pi m_e c^2 n^w_{e^\pm} R_s^2 \Gamma_w c$, where $n^w_{e^\pm}$ is the total number density of pairs in the wind region. The spindown energy from the pulsar sources the energy deposition on the wind region such that $L_w \approx L_{\rm sd}$ and thus
\begin{equation}
\Gamma_w = \frac{L_{\rm sd}}{4 \pi m_e c^2 n^w_{e^\pm} R_s^2 c} \,.
\end{equation}
The radius of the TS can be estimated by balancing the pressure in the nebula $P_{\rm neb}$ and the ram pressure $P_{\rm ram}$. Assuming a relativistic gas, the total pressure in the nebula $P_{\rm neb} = E_{\rm neb}^{\rm tot}/(3 V)$, where $E_{\rm neb}^{\rm tot}$ is the total energy in the nebula and $V \sim (4/3)\pi (R^3-R_s^3)$. The total energy in the nebula is composed of the non-thermal, thermal, magnetic energies and the energy due to $e^+-e^-$ pairs ($E_{\rm part}$). We thus have, $E_{\rm neb}^{\rm tot} = f_{\rm trap} (t) (E_{\rm nth} + E_{\rm th}) + E_B + E_{\rm part}$, where $f_{\rm trap} (t) = 1/\sqrt{1+(t/t_{\rm diff}^{\rm neb})^2}$, accounts for the fraction of non-thermal and thermal radiation that is trapped in the nebula. For $t \ll t_{\rm diff}^{\rm neb}$, that is, at the initial stage when typical nebular diffusion timescales are not relevant, the whole radiation is trapped in the nebula. For $t \gg t_{\rm diff}^{\rm neb}$, the fraction falls off as $\sim 1/t$ since nebular diffusion becomes efficient. The particle energy in the nebula can be estimated as $E_{\rm part} = (m_e c^2) \int d\gamma_e\ dN_e/d\gamma_e$ where $dN_e/d\gamma_e$ is the resulting spectrum after transport of the injected spectrum.

The ram pressure for a cold ultra-relativistic wind can be estimated by $P_{\rm ram} \simeq F_E/c$, where the energy flux $F_E = L_w/(4 \pi R_s^2)$. Balancing $P_{\rm neb}$ and $P_{\rm ram}$ we have
\begin{equation}
R_s^3 + \bigg( \frac{E_{\rm tot}c}{L_w} \bigg) R_s^2 - R^3 = 0\,,
\end{equation}
\begin{figure}
\centering
\includegraphics[width=0.49\textwidth]{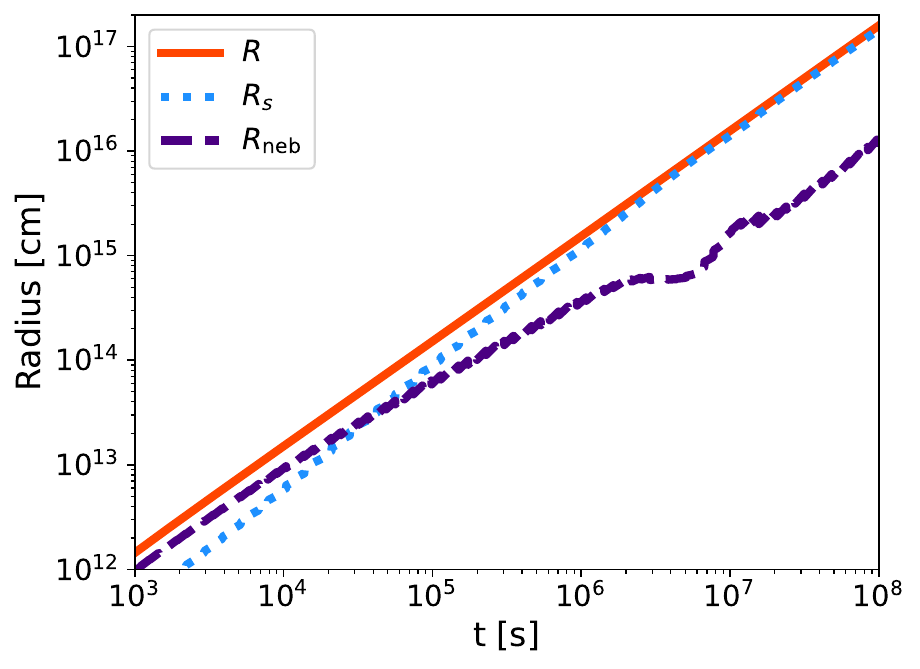}
\includegraphics[width=0.49\textwidth]{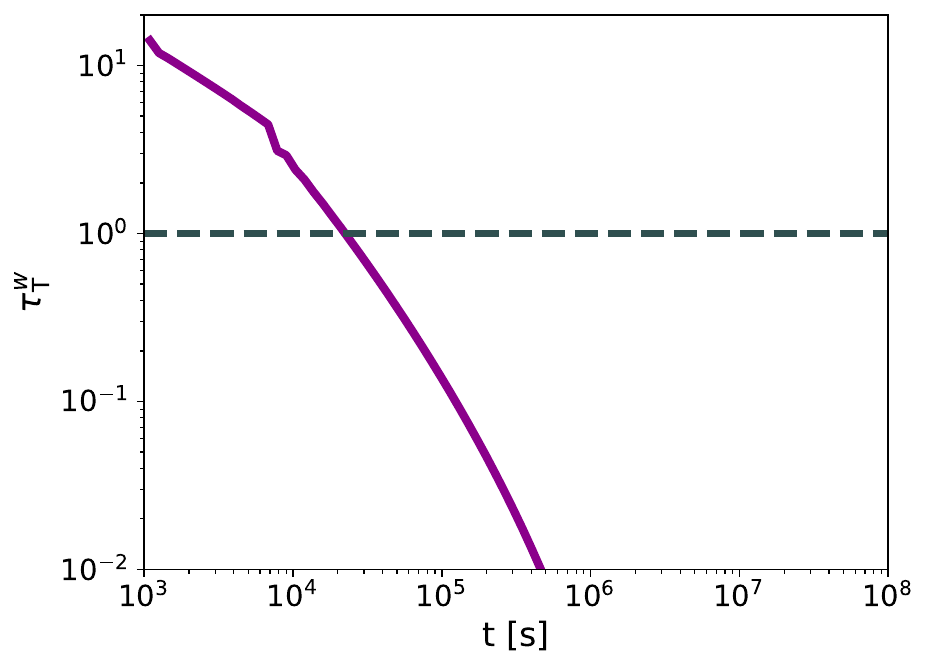}
\caption{\label{fig:rshock}\emph{Left: }Time evolution of $R_s$, $R$, and the width of the nebula $R - R_s$. \emph{Right: }Time evolution of the Thomson optical depth in the wind region. Note that $\tau_T^W (t \sim t_{\rm sd}) \ll 1$, therefore the TS is collisionless and not radiation mediated at late times.
}
\end{figure}
We solve the above equation to obtain $R_s$ such that $R_s \in \mathcal{R}$ and $R_s >0$, shown in Figure~\ref{fig:rshock} \emph{(left)}. Next, we evaluate the total number density of pairs in the wind region, that is, $n_{e^\pm}^w = N_{e^\pm}^w/V_w$, where the volume of the wind region $V_w \approx (4/3) \pi R_s^3$ and $N_{e^\pm}^w$ is the total number of pairs in the wind region. We assume that the density of pair creation rate through secondary two photon pair annihilation $\gamma \gamma \rightarrow e^+e^-$, is the same in the wind and the nebular region, that is, $\dot{n}^w_+ = \dot{n}^{\rm neb}_+$. This can then be used to estimate the pair creation rate in the wind region as
\begin{equation}
\label{eq:wind_paircreate}
\dot{N}^w_+ = \dot{N}^{\rm neb}_+ \bigg( \frac{R_s^3}{R^3 - R_s^3} \bigg)\,,
\end{equation}
where we assume that $\dot{N}^{\rm neb}_+$ is the pair creation rate from the secondary two photon pair annihilation $\gamma \gamma \rightarrow e^+e^-$. Similar to estimating the total number of pairs in the nebula ($N_\pm$), the total number of pairs in the wind region ($N_{e^\pm}^w$) can be estimated using Equation~\eqref{eq:pairs_neb} where we replace $\dot{N}^c_{e^\pm}$ with $\dot{N}^{\rm neb}_+$, $t_{e^\pm}^{\rm ann} \approx 1/\big( (3/16)\sigma_T n^w_{e^\pm} \big)$ and $t_{\rm dyn}^w \approx R_s/c$ and $V$ replaced by $V_w$. The Thomson optical depth in the wind region can be defined as $\tau_T^w \sim n_{e^\pm}^w \sigma_T R_s$. We show the time evolution of $\tau_T^w$ in Figure~\ref{fig:rshock} \emph{(right)} and note that $\tau_T^w \ll 1$ for $t \gtrsim 10^5$ s. This is crucial for a few assumptions we make in our formalism. Firstly, the assumption $\dot{n}^w_+ = \dot{n}^{\rm neb}_+$ becomes more reasonable for $\tau_T^w \ll 1$, so for the timescales relevant to us the assumption holds quite well. Secondly, for estimating $V_w$ we assume a volume filling factor instead of assuming a relativistic shell with volume swept up as $4 \pi R_s^2 c \delta t$. This is also reasonable since the pulsar wind fills a region enclosed by $R_s$, that is, the region with radius $r$ is filled where, $R_{\rm lc} < r < R_s$. As a consequence of $V_w$, the scaling in Equation~\eqref{eq:wind_paircreate} is also well-suited for our purpose. Furthermore, since $\tau_T^w \ll 1$ for $t>t_{\rm sd}$, the TS is collisionless and not radiation mediated. This implies that our model for shock acceleration at the TS is well-suited. But even though the acceleration at TS does play a role in deciding the maximum energy for the CR protons at late times (see Figure~\ref{fig:epenergies}), the CR protons mostly escape the system and do not produce neutrinos efficiently (see Figure~\ref{fig:timescales}).

The bulk Lorentz factor for the wind is shown in the \emph{right} panel of Figure~\ref{fig:pairmult}. Most importantly, we assume $\Gamma_w = \gamma_{\rm e,br}$, that is, the bulk Lorentz factor for the wind decides the break in the injected $e^+ - e^-$ spectra. At initial times, $\Gamma_w \sim 100$, while at late times the wind becomes ultra-relativistic with $\Gamma_w \gtrsim 10^6$. This is similar to what was found in~\cite{Vurm:2021dgo} (see Figure 11 there). This is to be expected since at late times the secondary pair creation rate $\dot{N}^{\rm neb}_+$ drops, resulting in inefficient pair loading in the wind and hence a large $\Gamma_w$.
\section{Photons in the nebula and ejecta}
\label{appsec:photons_neb_ej}
This section focuses on aspects of photons in the nebular and ejecta regions. In particular, we present details about the heating due to the decay of $^{56}$Ni leading to injection of MeV gamma rays in the nebula, the nebular photons spectra relevant for neutrino production, and the attenuation factor of the photons in the ejecta.
\subsection{Heating due to decay of $^{56}$Ni}
\label{appsubsec:ni_heat}
Nickel ($^{56}$Ni) with a half life of $\sim 10$ days decays to cobalt (Co) with a half life of $\sim 100$ days, which finally decays to iron (Fe), which can have an impact on the opacity relevant for photons. The decay of $^{56}$Ni is computed in the quasi-steady state approximation using
\begin{align}
N_{\rm Ni} &= N_{0,\rm Ni}\ \exp{(-t/\tau_{\rm Ni})}\,,\nonumber\\
N_{\rm Co} &= N_{0,\rm Co}+N_{0,\rm Ni}\ \big(1 - \exp{(-t/\tau_{\rm Ni})}\big)\,,\\
N_{\rm Fe} &= N_{0,\rm Fe}+N_{\rm Co}\ \big(1 - \exp{(-t/\tau_{\rm Co})}\big)\,\nonumber,
\end{align}
where the half-lives of Ni and Co are given by $\tau_{\rm Ni} = 8.8$ days and $\tau_{\rm Co}= 111.3$ days respectively. The initial number fractions of Ni, Co, and Fe are given by $N_{0,\rm Ni}$, $N_{0,\rm Co}$, and $N_{0,\rm Fe}$. The decay of $^{56}$Ni deposits thermal energy in the ejecta. This is achieved in the following way that the decay of $^{56}$Ni produces MeV gamma-rays which subsequently heat up their surrounding electrons through Compton scattering and produce non-thermal electrons in the process. The non-thermal electrons produced, subsequently thermalize by Coulomb interactions with the ambient plasma and by collisional ionization and excitation of atoms and ions, thereby depositing their energy as heat in the ejecta. The rate of the injected energy can be estimated by
\begin{align}
\label{eq:nidecay}
Q_{\rm Ni}^{\rm heat} &= f_{\rm therm} L_{\rm Ni}^{\rm decay}\,,\nonumber\\
f_{\rm therm} &= 1 - \exp{(-\tau_{\gamma})}\,,\\
L_{\rm Ni}^{\rm decay} = 10^{43}\ {\rm erg\ s}^{-1}&\left( 6.45\ e^{t/\tau_{\rm Ni}} + 1.45\ e^{t/\tau_{\rm Co}} \right) \left( \frac{M_{\rm Ni}}{1 M_\odot} \right)\,\nonumber,
\end{align}
where the luminosity of MeV gamma-rays due to the decay of $^{56}$Ni is given by $L_{\rm Ni}^{\rm decay}$, the fraction of thermalized photons is denoted by $f_{\rm therm}$. The opacity of these resulting MeV gamma-rays is given by $\kappa_\gamma = 0.05\ {\rm cm}^{2}g^{-1}$ (note that this is slightly different from the value we use for the ejecta opacity which is $\kappa_{\rm ej} = 0.1\ {\rm cm}^{2}g^{-1}$) such that the optical depth $\tau_\gamma = 3 M_{\rm ej} \kappa_\gamma/\big( 4 \pi R^2 \Gamma_{\rm ej}^2 \big)$.
\subsection{Nebular photon spectra}
\label{appsubsec:neb_photons}
\begin{figure}
\centering
\includegraphics[width=0.5\textwidth]{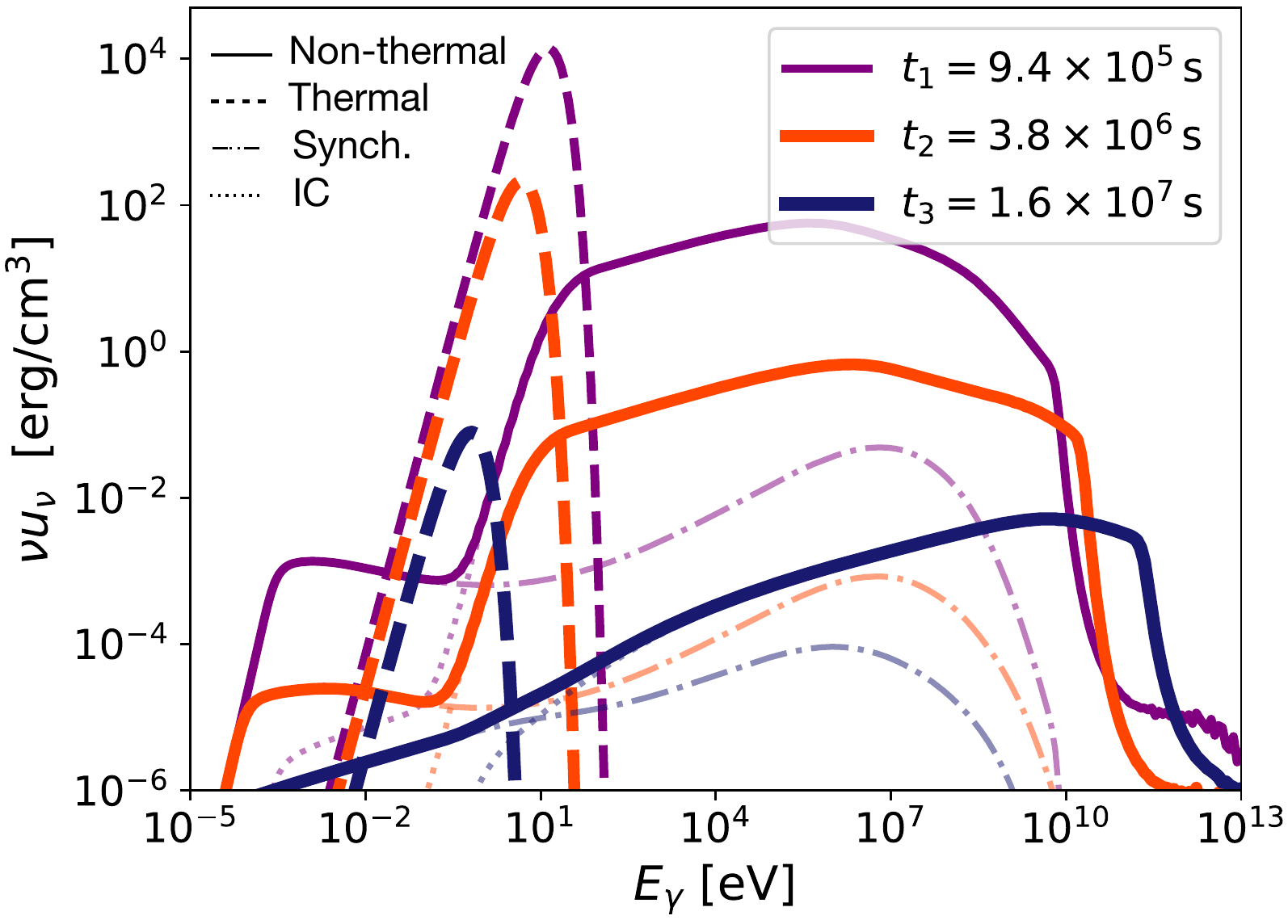}
\caption{\label{fig:photonspectra}Photon energy density spectrum ($\nu u_\nu$) of the non-thermal (solid) and thermal (dashed) photons in the nebula (comoving frame) after $\gamma \gamma$ and SSA attenuations, at different timesnaps $t_1 < t_{\rm sd}$, $t_2 \sim t_{\rm sd}$, and $t_3 > t_{\rm sd}$. The corresponding synchrotron (Synch.) and inverse Compton (IC) components are shown with thin dashed-dot and dotted lines respectively.
}
\end{figure}
In this subsection, we present the photon spectra in the nebula, which would be relevant for the neutrino production in the nebula through $p\gamma$ interactions (see~\citealt{Mukhopadhyay:2025tvz} for more details). Moreover, the photon spectra discussed here undergo attenuation in the ejecta (see Section~\ref{appsubsec:attn}) and result in the observational EM spectra discussed in Section~\ref{subsec:em_sig}.

The distribution of the energy density of non-thermal (solid) and thermal (dashed) photons is shown in Figure~\ref{fig:photonspectra} for three different timesnaps $t_1< t_{\rm sd}$ (purple), $t_2\sim t_{\rm sd}$ (orange), and $t_3 >t_{\rm sd}$ (dark blue). The differential energy density for the thermal photons assuming a blackbody spectrum with temperature $T_{\rm th} \sim \left( E_{\rm th}/\left( a_{\rm rad}\ (4/3) \pi R^3 \right) \right)^{1/4}$ is given by $u_\nu^{\rm th}=(8\pi h \nu^3/c^3)/\left[ \exp{\big(h\nu/(k_B T_{\rm th}) \big) - 1} \right]$, where $\nu$ is the frequency, $a_{\rm rad}$ is the radiation constant, $E_{\rm th}$ is the total energy in thermal photons (see Figure~\ref{fig:energies}), and $R$ is the radial distance of the nebula-ejecta boundary. With increase in time post the collapse the peak of the thermal photon energy distribution ($E_{\gamma,\rm th}$) shifts to lower energies. This can be understood by the fact that the thermal temperature decreases with time as a result of decrease in $E_{\rm th}$ (see Figure~\ref{fig:energies}) and increase in the radius of the nebula (see Figure~\ref{fig:rshock}, \emph{left} panel). The thermal temperature for the photons at the timesnaps can be estimated as $T_{\rm th}^{t_1} \approx 4 \times 10^4$ K ($E_{\rm th}^{t_1} \approx 1.4 \times 10^{50}$ erg and $R^{t_1} \approx 1.4 \times 10^{15}$ cm), $T_{\rm th}^{t_2} \approx 1.4 \times 10^4$ K ($E_{\rm th}^{t_2} \approx 7 \times 10^{49}$ erg and $R^{t_2} \approx 5.9 \times 10^{15}$) cm, and $T_{\rm th}^{t_3} \approx 1.9 \times 10^3$ K ($E_{\rm th}^{t_3} \approx 1.6 \times 10^{48}$ erg and $R^{t_3} \approx 2.5 \times 10^{16}$ cm). This implies $E_{\gamma,\rm th}^{t_1} \approx 9.7$ eV, $E_{\gamma,\rm th}^{t_2} \approx 3.4$ eV, and $E_{\gamma,\rm th}^{t_3} \approx 0.47$ eV for $t_1$, $t_2$, and $t_3$ respectively. The corresponding peak of the energy density in thermal photon is given by $u_{\nu, \rm th}^{t_1} \approx 1.2 \times 10^4\ {\rm erg/cm^3}$, $u_{\nu, \rm th}^{t_2} \approx 1.8 \times 10^2\ {\rm erg/cm^3}$, and $u_{\nu, \rm th}^{t_3} \approx 6.4\times 10^{-2}\ {\rm erg/cm^3}$. 

The magnetic energy density $u_B \sim \epsilon_B E_{\rm sd}/V$ at the different timesnaps $u_B^{t_1} \approx 2.2\ {\rm erg/cm^3}$, $u_B^{t_2} \approx 3.8 \times 10^{-2}\ {\rm erg/cm^3}$, and $u_B^{t_3} \approx 2.4 \times 10^{-4}\ {\rm erg/cm^3}$. Since $u_B \ll u_{\nu, \rm th}$ (see Figure~\ref{fig:energies}), the non-thermal photon energy density is dominated by inverse Compton (IC), with synchrotron emission contributing at energies lower than $E_{\gamma,\rm th}$ at the relevant time. The peak of the energy density for the non-thermal photon spectra can be estimated using $\nu u_{\nu, \rm IC} = L_{\rm nth}/(4 \pi R^2 c)$ such that $\nu u_{\nu, \rm IC}^{t_1} \approx 3.8 \times 10^{2}\ {\rm erg/cm^3}$, $\nu u_{\nu, \rm IC}^{t_2} \approx 6.6\ {\rm erg/cm^3}$, and $\nu u_{\nu, \rm IC}^{t_3} \approx 4.2 \times 10^{-2}\ {\rm erg/cm^3}$ (note that for $t_3$ the analytical estimate deviates slightly from the value obtained using numerics because in the late regimes the fast cooling approximation of pairs used in estimating $\nu u_{\nu,\rm IC}$ starts becoming less accurate). The decrease in the peak follows the decrease in $L_{\rm nth}$ as expected. The peak energy of the non-thermal photons can be estimated as $E_{\gamma,\rm nth} \sim \gamma_{e,\rm br}^2 E_{\gamma,\rm th}$. As seen from the figure peak shifts to higher energies with time since $E_{\gamma,\rm nth}^{t_1} \approx 8.7 \times 10^5$ eV ($\gamma_{e,\rm br}^{t_1} \sim 300$), $E_{\gamma,\rm nth}^{t_2} \approx 4.16 \times 10^6$ eV ($\gamma_{e,\rm br}^{t_2} \sim 1100$), and $E_{\gamma,\rm nth}^{t_3} \approx 1.9 \times 10^{10}$ eV ($\gamma_{e,\rm br}^{t_3} \sim 2 \times 10^5$).

At high-energies the photon spectra is suppressed by Klein-Nishina (KN) and $\gamma \gamma$ attenuation. The former can be estimated using $E_{\gamma, \rm KN} = \gamma_{e,\rm KN}^2 E_{\gamma,\rm th}$, where $\gamma_{e,\rm KN} \gtrsim (m_e c^2)/E_{\gamma,\rm th}$. Thus for the timesnaps $t_1$, $t_2$, and $t_3$ the KN suppression becomes relevant for $E_{\gamma, \rm KN}^{t_1} \approx 2.7 \times 10^{10}$ eV, $E_{\gamma, \rm KN}^{t_2} \approx 7.7 \times 10^{10}$ eV, and $E_{\gamma, \rm KN}^{t_3} \approx 5.6 \times 10^{11}$ eV respectively. For $\gamma \gamma$ cutoff above $E_{\gamma,\rm th}$ the target photon energy, in this case the energy in thermal photons decreases rapidly. In this scenario, the $\gamma \gamma$ cutoff can be estimated using $E_{\gamma \gamma,\rm cut} \approx (m_ec^2)/E_{\gamma,\rm th} \approx E_{\gamma,\rm KN}$.
\subsection{Attenuation in the nebula and ejecta}
\label{appsubsec:attn}
\begin{figure*}
\centering
\includegraphics[width=0.98\textwidth]{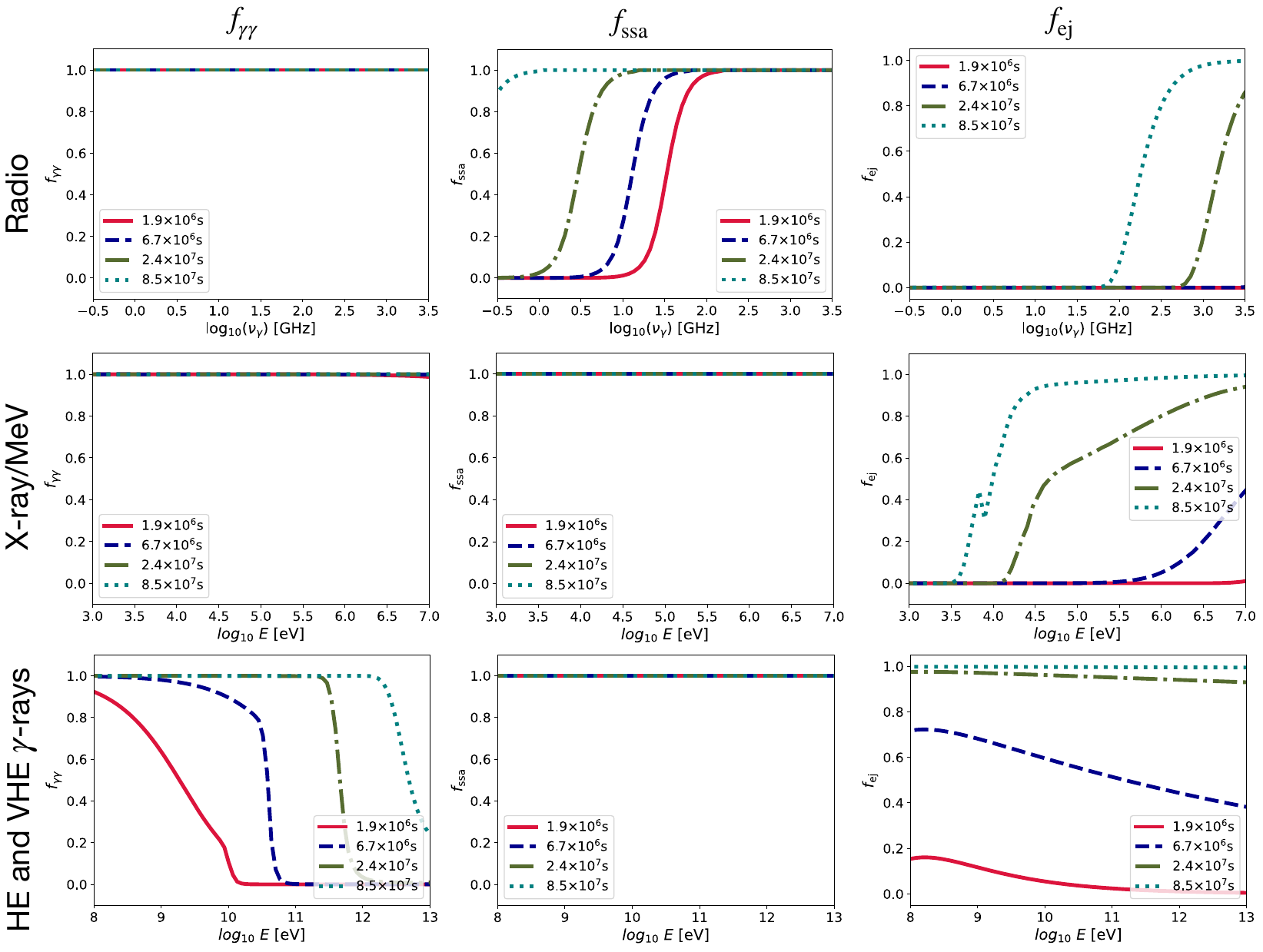}
\caption{\label{fig:attn} Attenuation coefficients at different time snaps for $\gamma \gamma$, SSA, and ejecta attenuations denoted by $f_{\gamma \gamma}$, $f_{\rm SSA}$, and $f_{\rm ej}$ respectively. The plots are shown for radio, X-ray/MeV, and high energy (HE) and very high energy (VHE) gamma-rays. The value of $f=1$ corresponds to $\tau=0$.
}
\end{figure*}
In this subsection we discuss the attenuation of the photon spectra in the nebula due to SSA and $\gamma \gamma$ processes, followed by the attenuation in the ejecta across all wavelengths due to free-free, bound-bound, bound-free, Compton and Bethe-Heitler processes (see~\citealt{Mukhopadhyay:2025tvz} for additional details). The observed EM spectra shown in Figure~\ref{fig:emspectra} is a result of the attenuation of the photons in the nebula and the ejecta regions.

The attenuation factors are defined in the following way $f_{\gamma\gamma}=(1-e^{-\tau_{\gamma\gamma}})/\tau_{\gamma\gamma}$, $f_{\rm ssa}=(1-e^{-\tau_{\rm ssa}})/\tau_{\rm ssa}$, and $f_{\rm ej}=e^{-\tau_{\rm ej}}$, where $f  \rightarrow 1$ corresponds to optical depth $\tau \rightarrow 0$. The attenuation factors across the radio, X-ray/MeV, and high-energy (HE) and very HE gamma-rays are shown in Figure~\ref{fig:attn} for different timesnaps. For the radio band the SSA optical depth and the free-free absorption in the ejecta highly suppress the emission. X-ray/MeV photons are suppressed mostly by the bound-free opacity of ejecta. Since the bound-free opacity is higher for lower energies, the ejecta becomes transparent to X-ray photons later than the MeV photons, explaining the rise of the MeV photon spectra at earlier times than the X-ray photon spectra. The MeV photons are scattered and reprocessed to lower energies through Compton scattering, which serves as the dominant attenuation channel.

Gamma-rays at GeV and TeV energies are mostly suppressed by the $\gamma \gamma$ attenuation in the nebula followed by Bethe-Heitler attenuation in the ejecta. The GeV photon spectra is observed earlier than the TeV photons because the $\gamma\gamma$ optical depth increases with increase in $E_\gamma$. Thus, even though the ejecta becomes transparent to GeV photons at $t \sim 10^7$ s post collapse, the $\gamma \gamma$ suppression completely suppresses TeV emission until very late times $t \sim 2.4 \times 10^7$ s.
\section{Hadronic processes in the nebula}
\label{appsec:hadronic_neb}
This section summarizes the hadronic processes in the nebula. In particular, the maximum energy achievable by the CR proton as a result of acceleration in the polar cap and TS regions and the associated timescales for acceleration and energy loss.
\subsection{Maximum CR proton energies}
\label{appsubsec:proton_energies}
\begin{figure}
\centering
\includegraphics[width=0.5\textwidth]{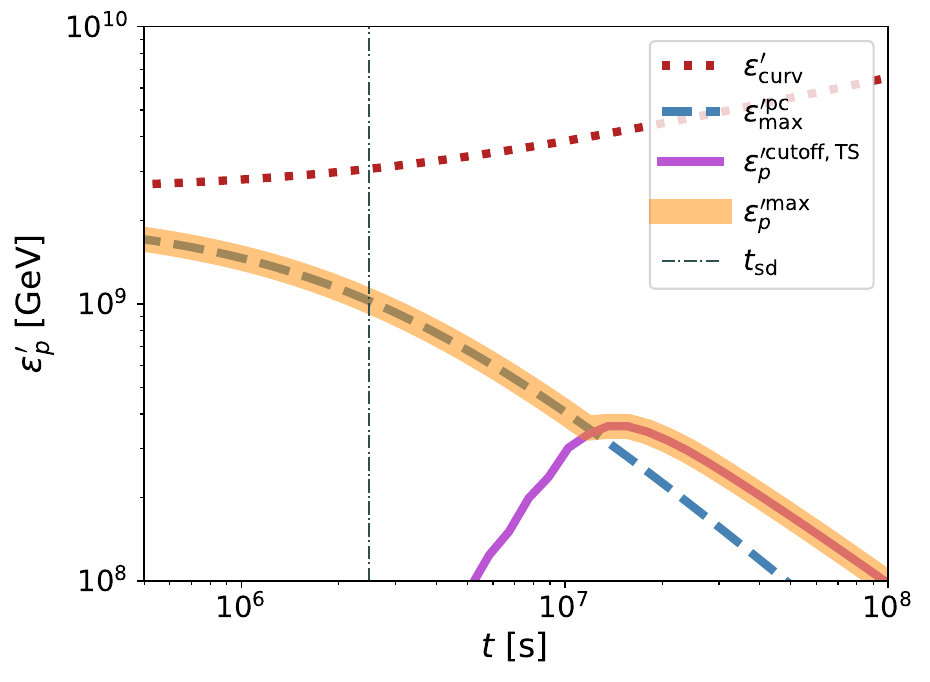}
\caption{\label{fig:epenergies} Time evolution of the relevant CR proton energies in different regions: the maximum energy that the protons can be accelerated to in the polar cap and TS regions are denoted by $\varepsilon_{\rm max}^{\prime \rm pc}$ (Equation~\ref{eq:pc}) and $\varepsilon_p^{\prime \rm cutoff, TS}$ respectively, the proton energy limited by curvature losses in the polar cap region is denoted by $\varepsilon^\prime_{\rm curv}$ (Equation~\ref{eq:curv}). The cut-off energy for the CR protons in a SLSNe ($\varepsilon_p^{\prime,\rm max}$) is shown as a shaded orange-yellow band.
}
\end{figure}
In this subsection we focus on evaluating the maximum energy to which CR protons can be accelerated in a millisecond magnetar powered SLSNe. We refer the reader to~\cite{Mukhopadhyay:2024ehs} (see Section 4 there) for a detailed discussion of the relevant energies and briefly discuss the results. Recall that in our setup there are two sites of particle acceleration -- the polar cap region of the magnetar and the termination shock (TS).

The maximum energy to which protons can be accelerated in the polar cap region $\varepsilon_{\rm max}^{\prime \rm pc}$ (dashed light blue), the energy cut off due to curvature losses $\varepsilon_{\rm curv}^{\prime}$ (dotted maroon), the maximum energy that protons can be accelerated to in the TS region $\varepsilon_{p}^{\prime \rm cutoff, TS}$ (solid light purple), and the maximum energy that the CR protons can be accelerated to $\varepsilon_p^{\prime \rm max}$ (shaded orange-yellow band) are shown as a functions of time post collapse in Figure~\ref{fig:epenergies}. We notice that given the choice of our fiducial parameters the acceleration across TS governs $\varepsilon_p^{\prime \rm max}$ at late times $t>10^7$ s. However, at such late times neutrino production is not efficient since the protons mostly escape from the nebula as evident from Figure~\ref{fig:timescales} (\emph{right} panel).

The curvature energy loss can be defined in the following way
\be
\label{eq:curv}
\varepsilon_{\rm curv}^\prime = \gamma_p m_p c^2 = \left[ \frac{3 m_p^4 c^8 B_d R_{\rm curv}^2}{2 Ze}\right]^{1/4} = 2.6 \times 10^{9}\ {\rm GeV}\ Z^{-1/4} \bigg( \frac{B_d}{5 \times 10^{13} {\rm G}} \bigg)^{1/4} \bigg( \frac{P_i}{4\ {\rm ms}} \bigg)^{1/2} \bigg( 1 + \frac{t}{t_{\rm sd}} \bigg)^{1/4}\,,
\ee
where $\gamma_p$ is the Lorentz factor of the protons and $R_{\rm curv}$ is the radius of curvature, which we assume to be $R_{\rm lc} = c/\Omega$, $\Omega = 2\pi/P$ is the angular velocity and the spin period $P(t) = P_i \big(1+t/t_{\rm sd} \big)^{1/2}$. This implies that for $t \ll t_{\rm sd}$, $\varepsilon_{\rm curv}^\prime$ is roughly constant and for $t \gg t_{\rm sd}$ it increases as $\sim t^{1/4}$. The acceleration of protons in the potential gap created in the polar cap region can be estimated as\footnote{Note the factor in front should be $2$ instead of $4$, which was a typo in our previous work~\cite{Mukhopadhyay:2024ehs}. This is because we define the magnetic dipole moment  of the magnetar as $\mu = B_{d} R_*^3$ instead of $B_{d} R_*^3/2$ used in the literature. Thus the potential gap created $\Phi_{\rm mag} = \Omega^2 \mu/c^2$.}
\be
\label{eq:pc}
\varepsilon^{\prime \rm pc}_{\rm max}= 2 \eta_{\rm gap} (Ze) B_d \left( \frac{\pi R_*}{c P} \right)^2 R_* = 2 \times 10^9\ {\rm GeV}\ Z \bigg( \frac{\eta_{\rm gap}}{0.1} \bigg)  \bigg( \frac{B_d}{5 \times 10^{13} {\rm G}} \bigg) \bigg( \frac{R_*}{10\ {\rm km}} \bigg) \bigg( \frac{P_i}{4\ {\rm ms}} \bigg)^{-2} \bigg( 1 + \frac{t}{t_{\rm sd}} \bigg)^{-1}\,,
\ee
where $\eta_{\rm gap}$ is the fraction of the total potential difference experienced by the charged particle. We use $\sim 10\%$ of the potential gap which can be reasonable due to screening effects from the surrounding plasma. For $t>t_{\rm sd}$, $\varepsilon^{\prime \rm pc}_{\rm max}$ falls off as $t^{-1}$ and therefore, $\varepsilon_p^{\prime \rm max}$ is limited by $\varepsilon^{\prime \rm pc}_{\rm max}$ for late times.

The maximum CR proton energy in the TS region is evaluated by balancing the acceleration timescale $t_{\rm acc}^\prime$ with the loss timescale $t_{\rm loss}^\prime$. We discuss and show the various different timescales in the following section.
\subsection{CR proton timescales for neutrino emission}
\label{appsubsec:timescales}
\begin{figure*}
\centering
\includegraphics[width=0.32\textwidth]{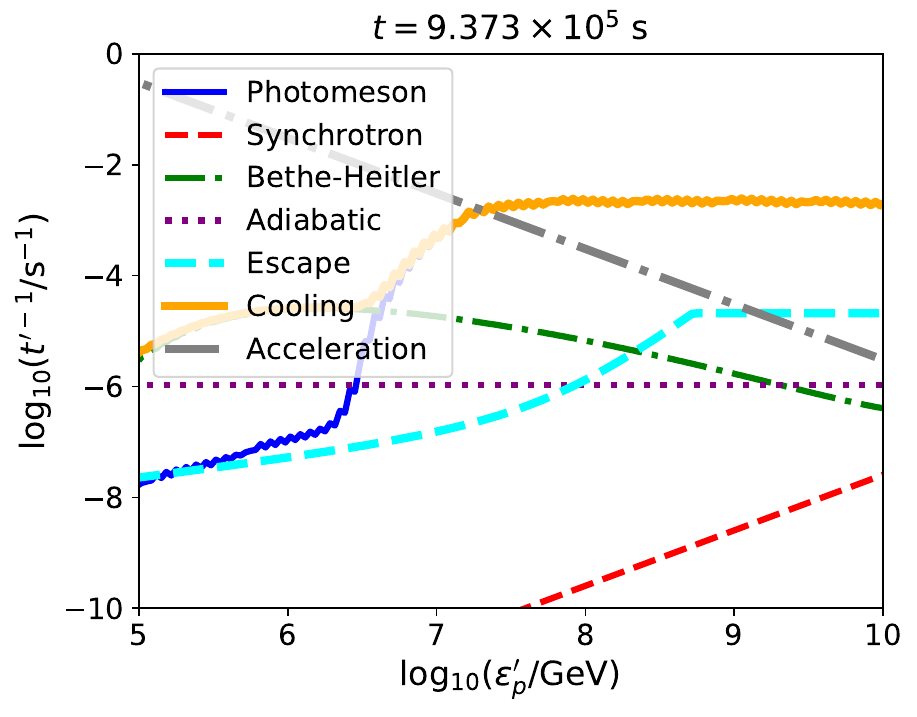}
\includegraphics[width=0.32\textwidth]{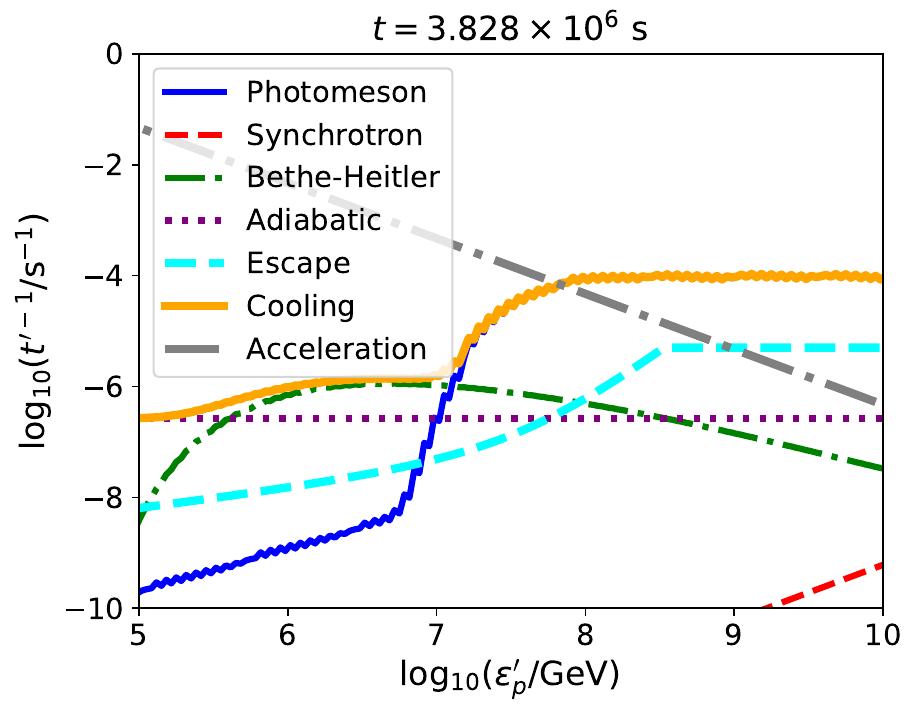}
\includegraphics[width=0.32\textwidth]{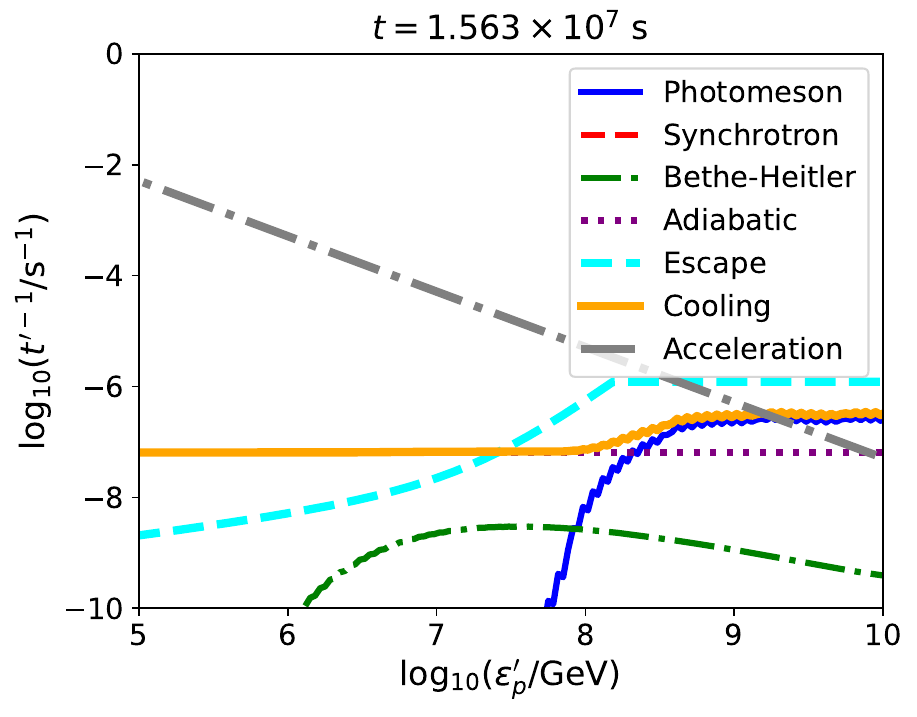}
\caption{\label{fig:timescales} Plots showing the rates of various acceleration and cooling processes with comoving proton energy $\varepsilon^\prime_p$ at various timesnaps. The total cooling rate $t_{\rm cool}^{\prime -1}$ is shown in solid yellow which includes contributions from $pp$ (not shown since it is negligible), $p\gamma$ (solid blue), proton synchrotron (dashed red), Bethe-Heitler (dot-dashed dark green), and dynamical/adiabatic (dotted pruple) cooling channels. The acceleration and escape rates for the protons are also shown in dot-dashed grey and dashed light blue respectively.
}
\end{figure*}
In this subsection we briefly discuss the relevant acceleration and loss timescales for the CR protons which are play a role in governing the the maximum energy of protons in the TS region and the efficiency of neutrino production. A more detailed description for the timescales can be found in~\cite{Mukhopadhyay:2024ehs} (see Section 4.3 there).

The acceleration timescale for the protons in the TS region can be estimated as $t_{\rm acc}^\prime \sim \eta_{\rm acc} \varepsilon_p^\prime/\big( e c B^\prime_{\rm neb} \big)$, where the comoving magnetic field in the nebula $B^\prime_{\rm neb} \approx \sqrt{2} \big( E_B/(R^2 v_{\rm ej} t) \big)$ and we choose the acceleration efficiency $\eta_{\rm acc} \sim 1$. The rate of energy loss for the protons is determined by their rate of escape ($t_{\rm esc}^{\prime -1}$) through diffusion or free-streaming and their rate of cooling ($t_{\rm cool}^{\prime -1}$). Furthermore, the cooling rate for the protons includes contributions from $pp$, $p\gamma$, synchrotron, inverse Compton, Bethe-Heitler, and dynamical processes, such that $t_{\rm cool}^{\prime -1} = t_{\rm pp}^{\prime -1} +  t_{\rm p\gamma}^{\prime -1} + t_{\rm sync}^{\prime -1} + t_{\rm BH}^{\prime -1} + t_{\rm dyn}^{\prime -1}$, where $t_{pp}^\prime , t_{\rm p\gamma}^\prime , t_{\rm sync}^\prime , t_{\rm BH}^\prime , t_{\rm dyn}^\prime $ denote $pp$, $p\gamma$, proton synchrotron (sync), Bethe-Heitler (BH), and dynamical/adiabatic (dyn) cooling timescales respectively. In Figure~\ref{fig:timescales} we show the rate of proton acceleration ($t_{\rm acc}^{\prime -1}$) (dot-dashed gray), cooling ($t_{\rm cool}^{\prime -1}$) (solid yellow) including the various channels, and escape ($t_{\rm esc}^{\prime -1}$) as a function of their comoving energy for different instances in time. Since the peak of neutrino emission occurs between $10^6 - 10^{6.5}$ s, we have chosen the timesnaps accordingly.

For $t<t_{\rm sd}$ (\emph{left} panel), protons with energy $\varepsilon^\prime_p \gtrsim 10^7$ GeV efficiently cool through $p\gamma$ interactions while $\varepsilon^\prime_p < 10^7$ GeV are cooled through BH processes. For $t\sim t_{\rm sd}$ (\emph{middle} panel), protons are accelerated efficiently upto $\varepsilon^\prime_p \lesssim 10^8$ GeV (however the maximum proton energy is still governed by the polar cap acceleration, see Figure~\ref{fig:epenergies}) and they efficiently cool through $p\gamma$ interactions. Furthermore, in this regime escape timescales are longer implying the $p\gamma$ cooling ideal for neutrino production through pions are efficient. Indeed we see that the neutrino fluence peaks around $t_{\rm sd} \sim 10^6 - 10^{6.5}$ s with the peak neutrino energy between $10^7 - 10^8$ GeV. The target photons relevant for neutrino production in this regime have energies $\sim 3 - 30$ eV. Given the thermal photons around $t_{\rm sd}$ have energy $E_{\gamma,\rm th}^{t_2} \sim 3.4$ eV, which implies that the thermal photons serve as the main target for neutrino production. At late times $t \gg t_{\rm sd}$ (\emph{right} panel) protons with energy $\varepsilon_p^\prime> 3\times 10^7$ GeV efficiently escape leading to a sharp decline in the neutrino production rate.
\section{Neutrino stacking searches using EM telescopes}
\label{appsec:nu_stack}
In this section we present the relevant details for evaluating prospects of stacking searches in the current and upcoming neutrino telescopes using optical observations of SLSNe in Rubin/Roman. The detection significance obtained as a result is shown in Figure~\ref{fig:nu} \emph{(right)} for different neutrino telescopes.
\subsection{Rate of SLSNe}
The redshift-dependent volumetric rate density of SLSNe is given by $\dot{\rho}(z) = \dot{\rho}_0 \big( \psi_z/\psi_0 \big)$, where $\dot{\rho}_0$ is the fiducial volumetric rate density at redshift $z = 0$, $\psi_z$ gives the evolution of the star formation rate with $z$, and $\psi_0 = \psi_{z = 0}$. We assume $\dot{\rho}_0 = 10^{-7}\ {\rm Mpc}^{-3}{\rm yr}^{-1}$ and $\psi_z$ is given by~\cite{Mukhopadhyay:2026wrv} (see Equation C4 there). Thus in a redshift bin $dz$ the observed SLSNe rate is given by
\be
\label{eq:rateslsne}
\frac{d \dot{N}}{dz} = \frac{dV_{\rm com}}{dz} \frac{\dot{\rho}(z)}{(1+z)} = 4 \pi \frac{c\ d_{\rm com}^2(z)}{H(z)} \frac{\dot{\rho}(z)}{(1+z)}\,,
\ee
where the all-sky comoving volume element is given by $dV_{\rm com}/dz = 4\pi\ dV_{\rm com}/(dz\ d\Omega)$ and we assume flat spacetime to obtain the simplification in the second equality, the redshift evolution of the Hubble expansion rate for flat $\Lambda$CDM cosmology is $H(z) = H_0\sqrt{\Omega_m (1+z)^3 + \Omega_\Lambda}$, where the matter and vacuum energy density fraction is given by $\Omega_m = 0.3$ and $\Omega_\Lambda = 0.7$ respectively. The comoving distance $d_{\rm com}$ is related to the luminosity distance $d_L$ using $d_L = (1+z) d_{\rm com}$. It is easy to see from above that the total number of SLSNe given a maximum redshift $z_{\rm max}$ and an observation time $T_{\rm obs}$ can be evaluated using $N_{\rm SLSNe} (T_{\rm obs}, z_{\rm max}) = T_{\rm obs}\int_{0}^{z_{\rm max}} dz\ \big( d\dot{N}/dz \big)$, where we have for $T_{\rm obs} = 10$ years and $z_{\rm max} = 1$, $N_{\rm SLSNe} \approx 4 \times 10^5$.
\subsection{Neutrino signal events from a population of SLSNe}
The observed differential neutrino number flux from a SLSNe at luminosity distance $d_L$ is defined as
\be
\label{eq:phinuobs}
\phi_{\nu}^{\rm obs} (z, t^{\rm obs}, E_\nu^{\rm obs}) = \frac{\phi_{\nu}^{\rm src}}{4 \pi d_L(z)^2}  = \frac{1}{4 \pi d_L(z)^2} \bigg(\frac{d^2 \dot{N}_\nu^{\rm src}}{dE_\nu^{\rm src} dt^{\rm src}}\bigg)\,
\ee
where the rate of neutrino production at the SLSNe (source) is given by $\phi_{\nu}^{\rm src}$ and is defined as term in the parenthesis in the second equality above. The source and observer frame neutrino energy is given by $E_{\nu}^{\rm src}$ and $E_{\nu}^{\rm obs}$ respectively, such that $E_{\nu}^{\rm src} = (1+z)E_{\nu}^{\rm obs}$. Similarly the source frame and the observer frame time interval is related by $dt^{\rm obs} = (1+z) dt^{\rm src}$. Using Equation~\eqref{eq:phinuobs}, the time-integrated observed neutrino number fluence is defined as
\be
\label{eq:jnuobs}
J_\nu^{\rm obs}(z, E_\nu^{\rm obs}) = \int dt^{\rm obs}\  \phi_{\nu}^{\rm obs} (z, t^{\rm obs}, E_\nu^{\rm obs}) = \frac{(1+z)}{4 \pi d_L(z)^2} \int_{t^{\rm src}_{\rm min}}^{{t^{\rm src}_{\rm max}}} dt^{\rm src} \left. \bigg(\frac{d^2 \dot{N}_\nu^{\rm src}}{dE_\nu^{\rm src} dt^{\rm src}}\bigg)\right|_{E_{\nu}^{\rm src} = (1+z) E_{\nu}^{\rm obs}}\,.
\ee
For a nearby source like a SLSNe at $d_L = 100$ Mpc, we can assume that the redshift $z \approx 0$ which simplifies the above expression and allows us to drop the superscripts. Hence in Figure~\ref{fig:nu} \emph{(left)} we show $E_\nu^2 J_\nu$ as the neutrino fluence from a SLSNe at $100$ Mpc ignoring the superscripts. We also choose $t^{\rm src}_{\rm min}$ and $t^{\rm src}_{\rm max}$ as half-decades in time from $10^5$ s to $10^{7.5}$ s to show the time-integrated neutrino fluences. The all-flavor effective areas ($A_{\rm eff}(E_\nu)$) for RNO-G, IceCube-Gen2 Radio, GRAND, and BEACON are obtained from~\cite{Kotera:2025jca} (see Figure 2 upper panel and references therein). Additionally, we also consider KM3NeT~\citep{KM3Net:2016zxf}, a typical $10\ {\rm km}^3$ telescope in the form of HUNT~\citep{Chen:2026jdx} or TRIDENT~\citep{TRIDENT:2022hql}, and NEON~\citep{Zhang:2024slv}. Given that both HUNT and TRIDENT will be water-based and have similar characteristics to KM3NeT, we use the KM3NeT effective area scaled by the detector volume ($\sim 10^{2/3}$) as an approximation for HUNT's and TRIDENT's effective area. 

Since the neutrino emission from SLSNe peak roughly a few tens of days post the collapse, the instantaneous effective area is not suitable and hence the day-averaged effective area needs to be used. While using the day-averaged effective area it is important to include the field of view (FOV) of the relevant detector since the day-averaged FOV is different from the instantaneous FOV for detectors not at the Pole. In our analysis, we consistently use the day-averaged effective area including the FOV. For IceCube we choose the all-flavor effective areas corresponding to the high-energy starting events (HESE)~\citep{IceCube:2020wum}. This choice is reasonable since the typical neutrino energy from SLSNe have $E_\nu > 10^5$ GeV. Finally, for IceCube-Gen2 we scale the effective area used for IceCube by a factor of $10^{2/3}$ due to the volume scaling of Gen2 with regards to current IceCube.

The total number of detected neutrino signal events from a single SLSNe at a redshift $z$ can be computed as $N_{\rm sig} (z) = \int dE_\nu^{\rm obs}\ J_{\nu}^{\rm obs} (z, E_{\nu}^{\rm obs}) A_{\rm eff}(E_{\nu}^{\rm obs})$. With all the ingredients discussed above (Equations~\ref{eq:rateslsne} and~\ref{eq:jnuobs}), the total number of stacked neutrino signal events that can be obtained from all SLSNe within a maximum redshift $z_{\rm maz}$ and observation time $T_{\rm obs}$ is defined as
\be
\label{eq:ntotsig}
N_{\rm sig}^{\rm tot}\big(T_{\rm obs}, z_{\rm max}\big) = T_{\rm obs} \int_{0}^{z_{\rm max}} dz\ \frac{d\dot{N}}{dz} \int_{E_\nu^{\rm min}}^{E_\nu^{\rm max}} dE_\nu^{\rm obs}\  J_{\nu}^{\rm obs} (z, E_{\nu}^{\rm obs}) A_{\rm eff}(E_{\nu}^{\rm obs})\,.
\ee
We choose the maximum redshift as $z_{\rm max} = 1$. This is typical of current and upcoming surveys like Rubin LSST, which can easily observe SLSNe for $z \sim 1$. Note that increasing $z$ does not necessarily guarantee an increase in the detection significance since increasing $z$ leads to an increase in $N_{\rm tot}^{\rm bkg}$ (Equation~\ref{eq:ntotbkg}) since $N_{\rm SLSNe}$ increases but $N_{\rm tot}^{\rm sig}$ may not increase comparatively since the contribution from far away sources become lesser as $d_L(z)$ increases. As evident from Figure~\ref{fig:nu} \emph{(left)}, the high energy neutrinos from SLSNe peak for $E_\nu > 10^5$ GeV, so we choose $E_{\nu}^{\rm min}$ and $E_{\nu}^{\rm max}$ as $10^5$ GeV and $10^{12}$ GeV respectively. This greatly helps in reducing atmospheric backgrounds.
\subsection{Background events}
The diffuse astrophysical neutrino flux that would serve as background is uncertain for $E_\nu > 10^6$ GeV. However, we provide an estimate of backgrounds by considering the latest background flux template from IceCube. In a stacking search the neutrino detector will also collect background events besides collecting signal events. We consider conventional, prompt, and astrophysical backgrounds for this work, that is, $\phi_\nu^{\rm bkg,obs} (E_\nu) = \phi_\nu^{\rm bkg, conv} + \phi_\nu^{\rm bkg, prompt} +  \phi_\nu^{\rm bkg, astro}$. We estimate the conventional atmospheric neutrino flux using the model described in~\cite{Honda:2006qj}, while the prompt atmospheric background is estimated from~\cite{Enberg:2008te}. The diffuse astrophysical background is estimated using the broken power-law fit to the IceCube data~\citep{IceCube:2024fxo}. This gives the diffuse isotropic astrophysical flux as $\phi_\nu^{\rm bkg, astro} (E_\nu) = \phi_0 \big( E_{\rm break}/100\ {\rm TeV}\big)^{-\gamma_2} \big( E_\nu/E_{\rm break}\big)^{-\gamma_2}$, where the all-flavor normalization $\phi_0 = 5.1 \times 10^{-17}\ {\rm GeV^{-1}cm^{-2}s^{-1}sr^{-1}}$, the spectral index $\gamma_2 = 2.52$, and the energy break $E_{\rm break} = 22.9$ TeV. Note that we just use the $E_\nu > E_{\rm break}$ branch of the broken power law since the SLSNe neutrino emission is dominated by neutrino energies $>10^5$ GeV. It is important to note that the IceCube analysis is dominated by neutrino flux below $100$ TeV. The extrapolation to $E_\nu > 10^7$ is optimistic since there could be additional  astrophysical sources at higher energies. Given we work with energies of 100 TeV and above, even though we consider the atmospheric fluxes, the background for $E_\nu > 10^5$ GeV is completely dominated by the astrophysical background, such that, $\phi_\nu^{\rm bkg,obs} (E_\nu) = \phi_\nu^{\rm bkg, astro} (E_\nu)$.

Any SLSNe detected by an EM telescope, that is subsequently used for performing a stacking search to look for neutrino events will contribute to the number of background events $N_{\rm bkg}$ that is collected. Given the isotropic background rate, the contribution will depend on the angular error associated ($\Delta\Omega_{\rm err}$) and the time window across which the search is performed ($\delta t_{\rm search}$). The former depends on the error associated with the localization of the source given the EM and neutrino detectors, that is, $\Delta \Omega_{\rm err} = {\rm max} \left[ \Delta \Omega_{\rm err}^{\rm EM}, \Delta \Omega_{\rm err}^{\nu} \right]$, while the latter depends on when the neutrino emission from the source in this case SLSNe peaks, that is, what would be the optimal time-window for the search. In general $\Delta \Omega_{\rm err}^{\rm EM}$ is independent of distance to the source while $\Delta \Omega_{\rm err}^{\nu}$ depends on the topology of the neutrino event in the detectors like tracks versus cascades. However, for EM telescopes, typical source localizations are at the level of arcseconds compared to a few degree localization from the neutrino telescopes, that is, $\Delta \Omega_{\rm err}^{\rm EM} \ll \Delta \Omega_{\rm err}^{\nu}$ and thus we can assume $\Delta \Omega_{\rm err} \approx \Delta \Omega_{\rm err}^\nu$. Thus for any observed SLSNe used for stacking, the corresponding background accumulated is $N_{\rm bkg} = \delta t_{\rm search} \Delta \Omega_{\rm err} \int dE_\nu^{\rm obs}\ \phi_\nu^{\rm bkg,obs} (E_\nu^{\rm obs}) A_{\rm eff}(E_\nu^{\rm obs})$. Note that the smaller the values of $\Delta \Omega_{\rm err}$ and $\delta t_{\rm search}$, the lower the accumulated backgrounds in stacking searches. Similar to Equation~\eqref{eq:ntotsig}, the total number of background neutrino events accumulated as a result of the stacking search can be defined as
\be
\label{eq:ntotbkg}
N_{\rm bkg}^{\rm tot} \big(T_{\rm obs}, z_{\rm max}\big) = N_{\rm SLSNe} N_{\rm bkg} = T_{\rm obs} \bigg( \int_{0}^{z_{\rm max}} dz\ \frac{d\dot{N}}{dz} \bigg) \bigg( \delta t_{\rm search} \Delta \Omega_{\rm err} \int_{E_\nu^{\rm min}}^{E_\nu^{\rm max}} dE_\nu^{\rm obs}\ \phi_\nu^{\rm bkg,obs} (E_\nu^{\rm obs}) A_{\rm eff}(E_\nu^{\rm obs}) \bigg)\,.
\ee
Note that since $\phi_\nu^{\rm bkg,obs}$ is independent of the redshift, the two integrals separate out. Furthermore, given the neutrino emission peaks for $t \sim 10^6 - 10^7$ s, we choose the optimal time-window for search as $\delta t_{\rm search} \sim 10^{6.4}\ {\rm s} \approx 30$ days. The assumption we make here is that all samples in our catalog of SLSNe for the stacking search has neutrino emission characteristics similar to our fiducial case. This can be improved by considering a scan over the parameter space to get a slightly more realistic estimate. Lastly, we choose the angular error for in-ice detectors like IceCube and IceCube-Gen2 to be $\theta_{\rm err}^{\nu,\rm ice} \sim 1^\circ$ while that for the radio neutrino detectors to be $\theta_{\rm err}^{\nu,\rm rad} \sim 0.5^\circ$, and water-based detectors $\theta_{\rm err}^{\nu,\rm water} \sim 0.25^\circ$ which gives $\Delta \Omega_{\rm err}^{\nu, \rm ice} = 2 \pi \big( 1 - \cos \theta_{\rm err}^{\nu,\rm ice}\big) \approx 10^{-3}\ {\rm sr}$, $\Delta \Omega_{\rm err}^{\nu, \rm ice} \approx 2 \times 10^{-4}$ sr, and $\Delta \Omega_{\rm err}^{\nu, \rm water} \approx 6 \times 10^{-5}$ sr respectively. This is reasonable since in-ice detectors indeed have slightly poor angular resolution than air shower detectors, while water-based detectors have excellent angular resolutions.

With these assumptions and choosing $z_{\rm max} = 1$ and $T_{\rm obs} = 10$ years, we find the background rate in a typical radio detector like IceCube-Gen2 Radio would be $\sim 3\ {\rm yr^{-1}}$. The detection significance in the Gaussian limit can be defined as
\be
\sigma_{\rm det} (T_{\rm obs},z_{\rm max}) = \frac{N_{\rm sig}^{\rm tot}}{\sqrt{N_{\rm sig}^{\rm tot} + N_{\rm bkg}^{\rm tot}}}\,,
\ee
where $N_{\rm sig}^{\rm tot}$ and $N_{\rm bkg}^{\rm tot}$ are given by Equations~\eqref{eq:ntotsig} and~\eqref{eq:ntotbkg} respectively. We show the detection significance for various current, upcoming, and proposed ice, air, and water based detectors neutrino detectors in Figure~\ref{fig:nu} \emph{(right)}.
\end{document}